\documentclass[range]{ar2e}%%%Put [range] option here for references citations to
\usepackage{graphicx}
\usepackage{bm}
\usepackage{epsfig}
\usepackage{epstopdf}
\usepackage{bm}
\usepackage{epsfig}
\usepackage{graphics}
\usepackage{xspace}
\usepackage{pstricks}
\usepackage{amssymb}

\def\OMIT#1{}

\newcommand{\nn}{\nonumber}

\newcommand{\bea}{\begin{eqnarray}}
\newcommand{\eea}{\end{eqnarray}}

\newcommand{\be}{\begin{equation}}
\newcommand{\ee}{\end{equation}}

\newcommand{\bfl}{\begin{flushleft}}
\newcommand{\bfr}{\begin{flushright}}
\newcommand{\efl}{\end{flushleft}}
\newcommand{\efr}{\end{flushright}}

\newcommand{\beq}{\begin{equation}}
\newcommand{\eeq}{\end{equation}}
\newcommand{\beqa}{\begin{eqnarray}}
\newcommand{\eeqa}{\end{eqnarray}}

\newcommand{\sstw}{\sin^2\theta_W}

\newcommand{\noi}{\noindent}
\newcommand{\lsim}{\mathrel{\lower4pt\hbox{$\sim$}}
\hskip-12.5pt\raise1.6pt\hbox{$<$}\;}
\newcommand{\gsim}{\mathrel{\lower4pt\hbox{$\sim$}}
\hskip-11.5pt\raise1.6pt\hbox{$>$}\;}
\newcommand{\MS}{$\overline{\rm MS}$}
\newcommand{\mMS}{\overline{\rm MS}}
\newcommand{\zzero}{Z^0}

\newcounter{saveeqn}

\begin{document}

%\input epsf.tex    %<-If you need EPS figures to be
                   %  called in {figure} environment for PC
%\input epsf.def   %<-If you need EPS figures to be
                   %  called in {figure} environment for Macintosh

%\input psfig.sty

\jname{Annu. Rev. Nucl. Part. Sci.}
\jyear{2013}
\jvol{1}
\ARinfo{1056-8700/97/0610-00}

\title{Low-Energy Measurements of the Weak Mixing Angle}

\markboth{Kumar, Mantry, Marciano \& Souder}{Low-Energy Measurements of the Weak Mixing Angle}

\author{K.S. Kumar
\affiliation{Department of Physics, University of Massachusetts, Amherst, MA 01003\\ 
email: kkumar@physics.umass.edu}
Sonny Mantry
\affiliation{Northwestern University and Argonne National Laboratory}
W.J. Marciano
\affiliation{Brookhaven National Laboratory}
P.A. Souder
\affiliation{Syracuse University}
}


\begin{abstract}

We review the status of precision measurements of weak neutral current interactions,
mediated by the $Z^0$ boson, at $Q^2\ll M_Z^2$. They can be used to extract 
values for the weak mixing angle $\sin^2\theta_W$, a fundamental parameter of the 
$SU(2)_L\times U(1)_Y$ electroweak sector of the Standard Model. Apart from
providing a comprehensive test of the electroweak theory at the quantum loop level, 
such measurements allow indirect access to new physics effects at and beyond the TeV scale. 
After a theoretical introduction and a brief overview
of the three most precise low $Q^2$ weak mixing angle determinations, 
we describe the ongoing experimental program and prospects for
future more sensitive studies. We also compare sensitivities of planned and proposed 
measurements to physics beyond the Standard Model.

\end{abstract}

\maketitle

\section{INTRODUCTION}\label{intro}

\subsection{Historical Context}
\label{sec:history}

In 1961, Sheldon Glashow~\cite{glashow61} introduced an SU(2)$_L\times$U(1)$_Y$ symmetry that would form the basis for electroweak unification. In modern terminology, it contained four spin-1 vector bosons $W^+_\mu,W^0_\mu,W^-_\mu$ and $B^0_\mu$ along with two independent couplings, $g$ and $g^\prime$. Mixing gave rise to a massless photon and its orthogonal massive partner now known as the $\zzero$ boson:
\bea
A_\mu = B^0_\mu\cos\theta_W + W^0_\mu\sin\theta_W  \nn \\
Z_\mu=W^0_\mu\cos\theta_W - B^0_\mu\sin\theta_W. 
\eea
\noi That formalism marked the birth of the weak mixing angle, $\theta_W$, 
defined by $\tan\theta_W = g^\prime/g$ or in terms of the electromagnetic coupling 
$e=gg^\prime/\sqrt{g^2+g^{\prime2}}$: $\sin\theta_W = e/g$.
Particle masses were arbitrarily put in by hand and were unrelated to other parameters in the theory.

In 1967, Steven Weinberg~\cite{weinberg67} appended the Higgs 
mechanism~\cite{higgs64, Higgs:1964pj, Higgs:1966ev, Englert:1964et, Guralnik:1964eu} 
to SU(2)$_L$ $\times$U(1)$_Y$ electroweak gauge unification via a complex, spin-0 scalar doublet whose vacuum
expectation value spontaneously broke the gauge symmetry to U(1)$_{em}$ and gave rise to $W^\pm$ and 
$\zzero$ masses related by $m_W=m_Z\cos\theta_W$. 
It also led to a physical spin-0 Higgs boson with arbitrary mass, $m_H$. As an added bonus, the Higgs which 
originally accommodated charged lepton masses, later proved well-suited to include quark masses and 
mixing~\cite{bouchiat72}, including CP violation. Weinberg speculated that the theory might be renormalizable and that weak neutral current effects, mediated by the $\zzero$ boson, should be observed in neutrino scattering. 

In 1971, Gerhard 't Hooft~\cite{thooft71} proved renormalizability for gauge theories with spontaneous symmetry breaking and weak neutral currents were discovered in 1973~\cite{hasert73}. 
Together, they confirmed the consistency and basic ingredients of electroweak unification.

The combination, SU(2)$_L\times$U(1)$_Y$ gauge invariance + Higgs doublet + renormalizability led to natural relationships among the bare (unrenormalized) gauge boson masses and couplings~\cite{bollini80}
\beq
\sin^2\theta^0_W = e^2_0/g^2_0 = 1-m^{0^2}_W/m^{0^2}_Z .\label{eqfive}
\eeq
\noi Those relations are respected by the renormalized parameters, up to finite calculable radiative 
corrections~\cite{bollini80, marciano75}. Such corrections, discussed in Sec.~\ref{secloops}, 
test the theory at its quantum loop level and probe for potential ``new physics" effects.

By the mid-1970s, the basic features of SU(2)$_L\times$U(1)$_Y$ electroweak unification were nearly established. The quark model (including charm) and its associated strong SU(3)$_c$ color gauge interactions, quantum chromodynamics (QCD), were elegantly incorporated while weak neutral current effects continued to be observed at about the predicted rate. However, it was not clear that the model's specific weak neutral current 
interaction~\cite{weinberg67}
\beq
\frac{g}{\cos\theta_W} Z_\mu \bar f\gamma^\mu (T_{3f} -2Q_f\sin^2\theta_W- T_{3f}\gamma_5)f,\quad T_{3f} = \pm1/2 \label{eqsix}
\eeq
\noi was correct. In particular, that interaction implied a small degree of parity violation throughout low energy physics due to $\gamma-\zzero$ interference. Such effects were enhanced in atoms with a
large number of protons ($Z$) and neutrons ($N$) by the coherent weak charge that quantifies the  
$\zzero$-nucleus vector coupling~\cite{bouchiat74}
\beq 
Q_W (Z,N) = Z(1-4\sin^2\theta_W) -N .\label{eqseven}
\eeq
\noi Unfortunately, the early efforts to measure atomic parity violation (APV) 
using $^{209}$Bi failed to observe the expected effect~\cite{lewis77, baird77} and cast
some doubt on the specific form of Eqn.~(\ref{eqsix}). (Later studies with $^{209}$Bi, $^{205}$Tl and $^{133}$Cs 
observed parity violation at the expected level~\cite{barkov78, conti79, bucksbaum82, bouchiat82}).  

In 1978, the now classic 
E122 experiment at SLAC, led by Charles Prescott searched for a helicity dependence in the inclusive cross-section
$\sigma_{R(L)}$ for deep-inelastic scattering of longitudinal polarized electrons off unpolarized 
$^2$H~\cite{prescott78}. Defining a parity-violating asymmetry
\beq
A_{PV} \equiv \frac{\sigma_R -\sigma_L}{\sigma_R + \sigma_L}, \label{eqeight}
\eeq
\noi the measured value, $A_{PV}\simeq 1.5\times10^{-4}$, confirmed 
the predicted form in Eqn.~(\ref{eqsix}) and determined $\sstw$ with relatively good precision 
($\pm10\%$): $\sin^2\theta_W \simeq 0.22(2)$.

That milestone experiment established SU(3)$_C\times$SU(2)$_L\times$U(1)$_Y$ as the Standard Model (SM). 
In addition, the measured value of $\sin^2\theta_W$ lent support to grand unified theories 
(GUTS)~\cite{georgi74, Georgi:1974yf} such as SU(5), SO(10),$\dots$. Assuming a ``great desert"
{\it i.e.} no ``new physics"
between the SM and unification scale $m_X\simeq10^{15}$ GeV, those theories predicted at one loop ($\alpha = e^2/4\pi \simeq 1/137$)~\cite{marciano79}
\beq
\sin^2\theta_W \approx \frac{3}{8} \left[ 1-\frac{109\alpha}{18\pi} \ln \frac{m_X}{m_W} \right] \simeq 0.21\qquad ({\rm One~loop~minimal~SU(5)}) \label{eqten}
\eeq
\noi in accord with the SLAC E122 value. (That simplistic scheme was, however, later ruled out by its failure to 
accommodate complete coupling unification and proton decay ($p\to e^+\pi^0$) constraints. 
Nevertheless, GUTS continue to be an interesting
paradigm; still used to advance low energy supersymmetry~\cite{marciano82} as a unifying desert ingredient.)

\subsection{The Weak Mixing Angle and Quantum Corrections}
\label{secloops}

In the 1980s, it became clear that in order to rigorously test the SM and GUTS at the level of their quantum corrections, $\sin^2\theta_W$ as well as the other electroweak parameters in Eqn.~(\ref{eqfive}) ($m_W$, $m_Z$, 
$\alpha$ and $G_F=g^2/4\sqrt{2} m^2_W$) would have to be determined with very high 
precision~\cite{marciano81, Marciano:1983wwa}, 
${\cal{O}}(\pm0.1\%)$ or better. Some were already known orders of magnitude better than needed; 
currently~\cite{aoyama12, tishchenko12} $\alpha^{-1} = 137.035999173(35)$ and 
$G_F = 1.1663787(6) \times 10^{-5}$ GeV$^{-2}$. 
In the case of vector boson masses, great progress was later made at LEP and the Tevatron~\cite{beringer12}:
$m_Z = 91.1876(21)$ GeV and $m_W = 80.385(15)$ GeV.

For $\sin^2\theta_W$, an important issue was the requirement of a rigorous definition of the renormalized weak mixing angle for precision experimental extraction. At first, the on-shell definition~\cite{sirlin80, Marciano:1980pb}
\beq
\sin^2\theta_W \equiv 1-m^2_W/m^2_Z \label{eqfifteen}
\eeq
\noi proved popular. However, after the top quark mass was found to be 
large (currently accepted value is $m_t = 173.3(8)$ GeV), 
the on-shell definition was largely abandoned, because its use induced large misleading 
${\cal O}(\alpha m^2_t/m^2_W)$ radiative corrections to weak neutral current processes. 
Instead, at LEP, it became practice to employ an effective 
$\sin^2\theta^{eff}_W$ defined by the $\zzero\mu^+\mu^-$ coupling at the $\zzero$ pole. The 
only drawback was the complexity of finite renormalized counterterms required for non-$\zzero$ pole applications.

For computational convenience and comparison with GUT predictions, it was easier to employ the more theoretically motivated (but unphysical) \MS\  (modified minimal subtraction) prescription 
(originally introduced for QCD)~\cite{marciano81, marciano81proc}
\beq
\sstw (\mu)_{\mMS} = e^2(\mu)_{\mMS}/g^2(\mu)_{\mMS} \label{eqseventeen}
\eeq
\noi with an arbitrary sliding mass scale $\mu$. Numerically, it is related to $\sstw^{eff}$ used at LEP 
by~\cite{gambino94}
\beq
\sstw(m_Z)_{\mMS} = \sstw^{eff} - 0.00028 \label{eqeighteen}
\eeq
\noi making translation between the two schemes straightforward.

Currently, the two best determinations of $\sstw(m_Z)_{\mMS}$ come from the right-left $Z$ pole production asymmetry $A_{RL}$ at SLAC~\cite{abe00}
\beq
\sstw(m_Z)_{\mMS} = 0.23070(26)\qquad\qquad A_{RL} \label{eqnineteen}
\eeq
\noi and the $Z\to b\bar b$ forward-backward asymmetry $A_{FB}(b\bar b)$ measured at LEP1~\cite{abreu95} 
\beq
\sstw (m_Z)_{\mMS} = 0.23193(29).\qquad\qquad A_{FB}(b\bar b) \label{eqtwenty}
\eeq
\noi Unfortunately, they disagree by 3.2 sigma. 
Even the overall LEP1 average including lepton forward-backward asymmetries and $\tau$ polarization, $\sstw (m_Z)_{\mMS} = 0.23161(21)$,
is somewhat high compared to Eqn.~(\ref{eqnineteen}). Nevertheless, all $\zzero$ pole measurements 
are usually averaged to give
\beq
\sstw (m_Z)_{\mMS} = 0.23125(16) \qquad\qquad Z {\rm~pole~Ave.} \label{eqtwentytwo}
\eeq
\noi for comparison with other precision studies. The spread in the most precisely measured values of $\sstw(m_Z)_{\mMS}$ remains, however, somewhat troubling and needs to be resolved, as underscored by 
an example of their different implications discussed toward the end of this subsection.

The exquisite precision achieved in the measurements of $m_Z$, $m_W$ and $\sin^2\theta_W$ allows for 
important tests of the electroweak theory at the level of quantum loops. In the process of renormalization,
finite radiative corrections upset the natural relations of Eqn.~\ref{eqfive}. The fractional deviation has been
historically~\cite{sirlin80, Marciano:1980pb} called $\Delta r$, and is primarily due to fermion and boson 
vacuum polarizations (including those involving
top quarks and SM bosons that are heavier than the energy scales of various measurements), but could also receive contributions from other indirect effects of even higher mass scale
``new physics". Conventionally, three different quantities have been used to parametrize the deviations
in the finite radiative corrections from 
zero~\cite{sirlin80, Marciano:1980pb, marciano93, ferroglio02, Awramik:2003rn, Awramik:2006uz}, 
because of their distinctly different dependencies on $m_t$, $m_H$ and ``new physics":
\beq
\label{eqtwentythree}
\begin{array}{rclcl}
(\Delta r)^{\rm expt}  & = & 1- \left[\pi\alpha/\{\sqrt2G_F m^2_W (1-m^2_W/m^2_Z)\}\right]   & = & 0.0350(9) \\   
(\Delta r)^{\rm SM}  & = & 0.0364(3) + 3.4\times 10^{-3}\ln \left[m_H/126{\rm ~GeV}\right] & &  
\end{array}
\eeq
\beq
\label{eqtwentyfour}
\begin{array}{rclcl}
(\Delta \hat r)^{\rm expt} & = & 1-\left[2\sqrt2\pi\alpha/\{G_Fm^2_Z\sin^22\theta_W(m_Z)_{\mMS}\}\right] & = & 0.0598(5)  \\
(\Delta \hat r)^{\rm SM} & = & 0.0598(2) +  1.4\times 10^{-3}\ln \left[m_H/126{\rm ~GeV}\right] & & 
\end{array}
\eeq
\beq
\label{eqtwentyfive}
\begin{array}{rclcl}
(\Delta r_{\mMS})^{\rm expt} & = & 1-\left[\pi\alpha/\{\sqrt2 G_Fm^2_W\sstw(m_Z)_{\mMS}\}\right] & = & 0.0699(7)(4) \\
(\Delta r_{\mMS})^{\rm SM} & = & 0.0693(2) +  6.5\times 10^{-4}\ln \left[m_H/126{\rm ~GeV}\right] & & 
\end{array}
\eeq
\noi where the dependence of the first two corrections on $m_H$ provide sensitivity to it, 
while $\Delta r_{\mMS}$ has 
less dependence on $m_H$. We emphasize that the values on the right (on the first line of each equation) are
purely experimental determinations. 
Those on the second line of each equation incorporate detailed calculations of SM
loop corrections~\cite{ferroglio02, Awramik:2003rn, Awramik:2006uz, marciano04, Marciano:2000yj} 
(assuming no ``new physics") and using experimental measurements of $\alpha_{EM}$, $G_F$, 
$m_Z$ and $m_t$ as input. The theoretical predictions
are dominated  by a +7\% shift due to fermion vacuum polarization effects that lead to the 
running of $\alpha_{EM}$ from $\alpha(0) = 1/137$ to $\alpha(m_Z)_{\mMS}= 1/127.9$, but also
include important dependences on $m_H$ and $m_t$. 
Note also, that we have normalized the predictions at $m_H=126$ GeV, the tentative value of the new scalar 
resonance recently discovered at the Large Hadron Collider~\cite{aad12,:2012gu}.

Alternatively, one can obtain, from the first two corrections the predictions 
(Eqns.~\ref{eqtwentythree}--\ref{eqtwentyfour}):
$m_W = 80.362(6)$ GeV and $\sstw(m_Z)_{\mMS} = 0.23124(6)$ for $m_H = 126$ GeV, 
where the uncertainties are due to the errors in $m_t$ and hadronic effects.  (The errors are 
approximately doubled if one includes estimated uncertainties in uncalculated higher order 
effects~\cite{ferroglia12}).  
The agreement between those predictions and the corresponding  world averages of current measurements
constitutes a beautiful verification of the electroweak theory
at the quantum loop level and constrains many ``new physics" 
scenarios. 
If instead, one takes the world averages for $m_W$ and $\sstw$, one obtains 
$m_H  =  97^{+24}_{-20}$ GeV
in relatively good accord with the LHC finding but somewhat low, still leaving some room for ``new physics". 

The experimental determinations of $m_W$ and $\sin^2\theta_W$ also provide a direct probe of ``new physics" 
by testing the validity of the third $\Delta r_{\mMS}$ relation (Eqn.~\ref{eqtwentyfive}).
For example, taking $m_H\simeq 126$ GeV but allowing for $N_D$ heavy new chiral doublets ($N_D=4$ for a fourth
generation) via $S=N_D/6\pi$ or a heavy $W^{\ast\pm}$ excited $W$ boson leads, upon comparing with
experiment,  to~\cite{marciano90, marciano11}
\beq
\Delta r_{\mMS}(m_Z) = 0.0693(2)+0.0085S + \left(\frac{m_W}{m_{W^*}}\right)^2=0.0699(8) \label{eqthirty}
\eeq
\noi or
\bea
S=0.07(9) \to N_D\le4 \qquad ({\rm One-sided~ 95\%CL}), \nn \\
m_{W^*}>2.2 {\rm~TeV} \qquad ({\rm One-sided~ 95\%CL}). 
\eea
\noi Such constraints appear to tightly restrict ``new physics''. However, they are quite dependent on the $\zzero$ 
pole average $\sstw(m_Z)_{\mMS}$ employed as well as the overall error. If one instead uses the $A_{fB}(b\bar b)$ 
value in  Eqn.~\ref{eqtwenty}, it suggests $S\sim 0.4$ or $N_D \simeq 6$--7, more in keeping with 
dynamical symmetry breaking (technicolor) or 4th generation scenarios. Additionally, in Eqn.~\ref{eqtwentyfive}, the larger(smaller) 
error is due to the uncertainty in $\sin^2\theta_W$($m_W$). This
underscores the need for improved experimental determinations of $\sstw$, the topic of this review.

\subsection{The Weak Mixing Angle at $Q^2\ll M_Z^2$}
\label{seclowqsin}

What do low energy determinations of $\sstw$, the subject of this review, add to the above discussion? How do they complement the already precise $\zzero$ pole measurements?
Currently, there are 3 low $Q^2$ measurements of $\sstw$ at the $\pm1\%$ level or better. They will be reviewed in Sec.~\ref{secpast}, including details and caveats associated with each extraction. Here, we summarize the results
extrapolated to $m_Z$ scale for comparison with $\zzero$ pole 
measurements~\cite{Dzuba:2012kx, Anthony:2005pm, Zeller:2001hh}
\beqa
\sstw(m_Z)_{\mMS} & = & 0.2283(20) \qquad {\rm Cs~APV~at~}\langle Q\rangle \simeq2.4 {\rm~MeV} \label{eqthirtythree} \\
\sstw(m_Z)_{\mMS} & = & 0.2329(13) \qquad {\rm M\o ller}~A_{PV}~{\rm at}~ \langle Q\rangle \simeq160 {\rm~ MeV} \label{eqthirtyfour} \\
\sstw(m_Z)_{\mMS} & = & 0.2356(16) \qquad \nu_\mu N~{\rm at}~ \langle Q\rangle\simeq 5 {\rm~GeV} \label{eqthirtyfive} 
\eeqa
\noi Those values are not directly competitive with $\zzero$ pole results. Even the average of 
Eqns.~\ref{eqthirtythree}--\ref{eqthirtyfive}: $\sstw(m_Z)_{\mMS} = 0.2328(9)$ {\it i.e.} ${\cal{O}}(\pm0.4\%)$
lends little to the above discussion.
However, as we discuss in Sec.~\ref{secprogram}, 
future polarized electron scattering asymmetries at low $Q^2$ are expected to
reach similar precision to the best $\zzero$ pole measurements: ${\cal{O}}(\pm0.1-0.2\%)$. 
At that level, they may help resolve differences between the
SLAC and LEP1 results, or perhaps, as we shall discuss, 
find interesting new effects.

Apart from improved precision testing of the electroweak theory at the quantum loop level, 
low $Q^2$ measurements are sensitive to classes of ``new physics" effects to which $\zzero$ pole
measurements are insensitive. 
The measurements in Eqns.~\ref{eqthirtythree}--\ref{eqthirtyfive} can already be used to
constrain ``new physics'' such as $Z^\prime$ bosons or general 4-fermion contact interactions. Future
more precise experiments are expected to probe the 1--20 TeV scale, as described in Sec.~\ref{secnew}.
We also show how such experiments may be used to explore very weakly coupled low mass scale
``dark boson" effects.

In addition, the low $Q^2$ results already test the SM predicted 
running~\cite{marciano81, Marciano:1983wwa, marciano81proc, Czarnecki:1995fw, czarnecki98, czarnecki00} 
of $\sstw$ as a function of $Q^2$. The evolution of that quantity can be examined in the \MS\ framework of 
Eqn. (\ref{eqseventeen}) using the $e(\mu)$ and $g(\mu)$ beta functions. 
\begin{figure}[tb]
\includegraphics[height=4.6cm]{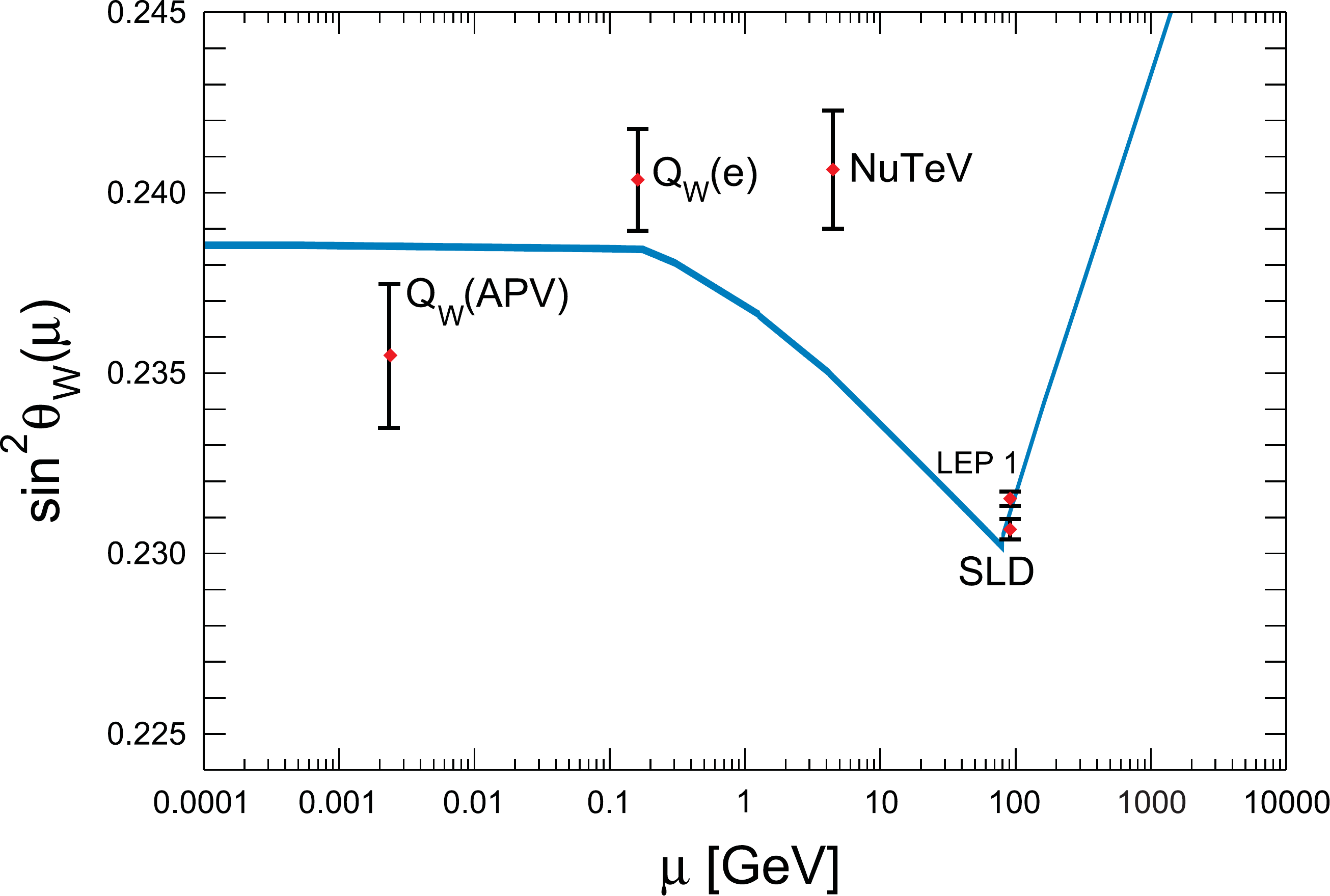}
\includegraphics[height=4.8cm, width=6.6cm, trim = 0mm 1mm 0mm 0mm]{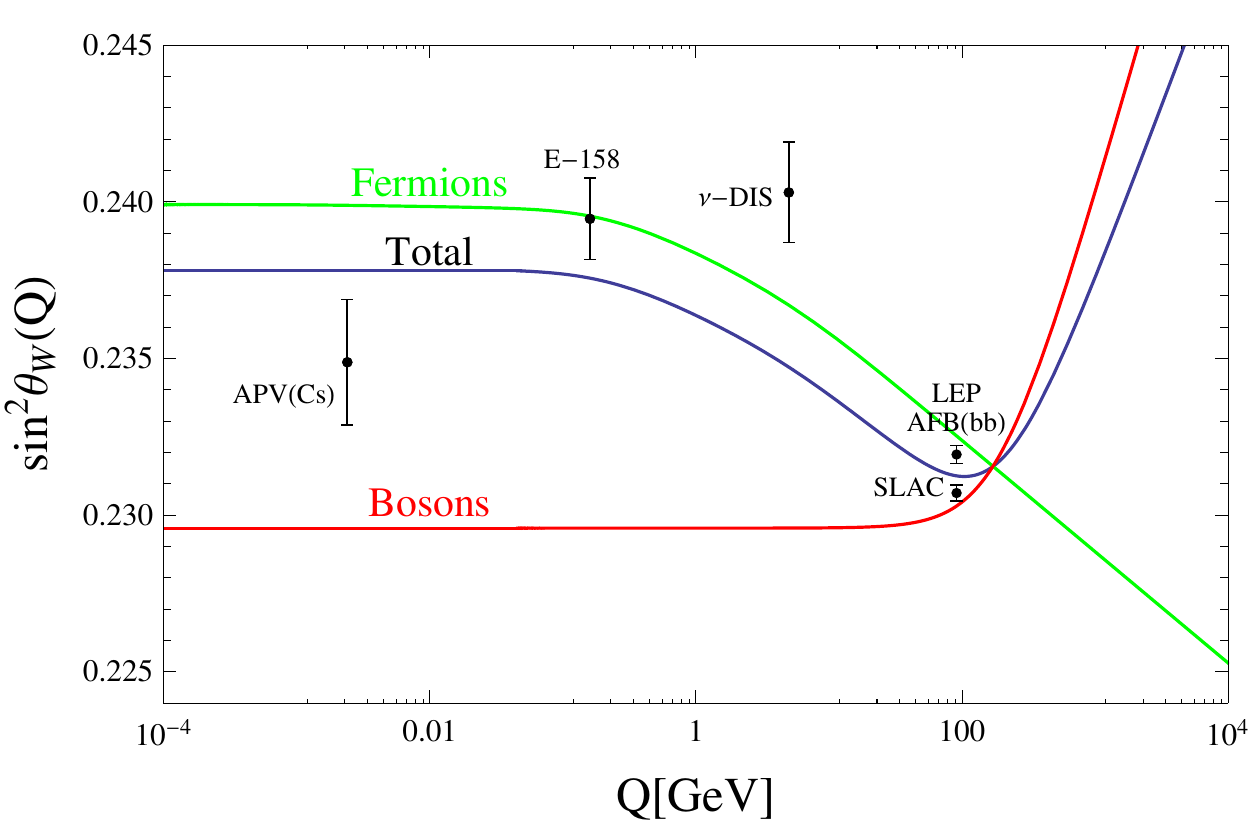}
\caption{$\sstw(\mu)_{\mMS}$ (left panel) from Ref.~\cite{beringer12} with updated APV result.
$\sstw(Q^2)$ (right panel), a one-loop calculation dominated by $\gamma-\zzero$ 
mixing~\cite{czarnecki00}. The red and green curves are the boson and fermion contributions respectively.}
\label{fig:run}
\end{figure}
The running is illustrated in Fig.~\ref{fig:run} along with the three measurements
with $\mu=\langle Q\rangle$ determined from each experiment's average momentum conditions~\cite{Erler:2004in}. 
The only drawback to that formalism is the unphysical \MS\ discontinuities
 at $\mu=\langle Q\rangle ={}$particle masses. To circumvent that feature, it is useful to define a 
 more physical running weak angle~\cite{Czarnecki:1995fw, czarnecki98, czarnecki00, Ferroglia:2003wa}
\beq
\label{thetaQ2}
\sstw(Q^2) = \kappa(Q^2)\sstw (m_Z)_{\mMS} \label{eqthirtysix}
\eeq
\noi where $\kappa(Q^2)$ incorporates perturbative $\gamma-Z$ mixing through vacuum polarization and other 
smaller corrections. This is illustrated in Fig.~\ref{fig:run} normalized such that $\kappa(Q^2=m^2_Z)\simeq1.000$
while $\kappa(0)$ turns out to be about 1.030. This 3\% variation is particularly important for some low
$Q^2$ polarized electron scattering asymmetries proportional to 
$1-4\sstw(Q^2)$ (examples discussed in Sec.~\ref{secprogram}) that are very sensitive to small
variations in $\sstw(Q^2)$. Indeed, the 3\%\ shift in $\sstw$ results in a roughly 40\%\ change
in $1-4\sstw$, ($0.075\to 0.046$ at $\langle Q\rangle\sim 0.1$ GeV).  

Of course, to test the running of $\sstw(Q^2)$ and try to unveil ``new physics'' requires confidence in the theoretical
 underpinnings of the various reactions studied. To that end, we examine in Sec.~\ref{sectheory} the status of
 several theoretical issues, including hadronic uncertainties in $\kappa(0)$ and $\gamma\zzero$ box diagrams.  
 However, we emphasize that each experimental measurement should be
compared with a calculation specific to the relevant experimental conditions including
complete one-loop and leading two-loop effects and estimates of hadronic contribution uncertainties.
The latter are under control for very low $Q^2$ and high $Q^2$ measurements
but may require more careful studies at intermediate $Q^2$.
We conclude with a brief sketch of other potential ways to measure $\sin^2\theta_W$ in Sec.~\ref{secother}, and 
provide a future perspective and outlook in Sec.~\ref{secconclude}.

\section{PAST MEASUREMENTS}
\label{secpast}

In the following section, we review the three most precise published measurements of 
$\sin^2\theta_W$ at $Q^2\ll M_Z^2$. The implications of the measurements for high and low scale dynamics will
be addressed in Sec.~\ref{secnew}. A comprehensive review of earlier experiments and associated
developments can be found in Ref.~\cite{Musolf:1993tb}; a very recent review addresses a broader class
of weak neutral current observables~\cite{erlersu}. 

\subsection{Atomic Parity Violation in Cesium}
\label{secAPV}

One of the classic precision techniques in the field is the measurement of 
parity violation in atoms, as mentioned in Sec.~1. The electron-nucleus weak neutral current 
interaction mediated by $\zzero$ exchange can be characterized by a new term in the Hamiltonian
with an overall strength $Q_WG_F$, where $G_F$ is the Fermi constant and 
the weak charge $Q_W$ was defined in Eqn.~\ref{eqseven}. 
The new interaction induces a parity-violating matrix element 
$Im(E1_{PNC})=Q_Wk_{PNC}/N,$
where $k_{PNC}$ is a quantity that can be computed from the atomic wavefunctions.

The experiment determines the ratio of $E1_{PNC}$
to a Stark mixing matrix element ${Im(E1_{PNC})}/{\beta}$,
where $\beta$ is the vector transition polarizability.
In 1997, the most precise result to date utilizing Cesium was measured~\cite{Wood:1997zq}:
${Im(E1_{PNC})}/{\beta}=1.5935(56)~{\rm mV/cm}$.
The value of $Q_W$ is obtained from
\begin{equation}
Q_W=\left(\frac{E1_{PNC}/\beta}{M_{hf}/\beta}\right)
\left(\frac{NM_{hf}}{k_{PNC}}\right)
\end{equation}
\noi where $\beta$ and $k_{PNC}$ were determined from atomic theory.
In 1999, a more 
precise value of $Q_W$ was extracted~\cite{Bennett:1999pd} based on two improvements: 
$M_{hf}/{\beta}$ was measured and $\beta$ was obtained
from a precise calculation of $M_{hf}$.
Second, the theoretical error in $k_{PNC}$ was evaluated by benchmarking the
  calculation with other measurable quantities, such as hyperfine levels.
With the improved data, $k_{PNC}=0.9065(36)\times10^{-11}ea_0$ was obtained.
The new analysis gave the result
$Q_W=-72.06~(28)_{\rm expt}~(34)_{\rm theor}$
that differed from the SM prediction by 2.3$\sigma$.

Over the past decade, several theoretical developments appeared to resolve the discrepancy with 
the SM (such as the inclusion of additional QED corrections and the Breit correction). The most detailed new
corrections emerged from a new, high-precision calculation~\cite{Porsev:2010de} that used
the coupled cluster approximation and included triple excitations in addition
to the single and double excitations that were treated in earlier
calculations. The result was $k_{PNC}=0.8906(26)\times10^{-11}ea_0$, 
which led to $Q_W({\rm Cs})^{\rm exp} = -73.16(28)(20)$
in excellent agreement with the SM expectation.  However, a recent reevaluation~\cite{Dzuba:2012kx}
 of some of the contributions has changed the result to
$k_{PNC}=0.8977(40)\times10^{-11}ea_0$ which is more consistent with earlier work~\cite{Dzuba:2002kx}. 
This leads to $Q_W = -72.58(43)$ and a value for $\sstw$
quoted in Eqn.~\ref{eqthirtyfive} of Sec.~1, when compared to the latest theoretical 
prediction, including a small Pauli blocking correction~\cite{Blunden:2012ty}. Indeed, one expects, including
updated electroweak corrections,
$Q_W({\rm Cs})^{\rm SM} = -73.24(5)$, a difference of 1.5$\sigma$ from experiment. 
As discussed in Sec.~\ref{secnew}, the result nevertheless continues to 
significantly constrain new TeV-scale lepton-quark interactions, complementing direct collider searches.

\subsection{SLAC E158}
\label{secE158}
After parity violation in neutral currents was observed by the 
SLAC E122 experiment as discussed in Sec.~1, 
the possibility was considered of measuring parity violation in electron-electron (M{\o}ller)
scattering. The value of $A_{PV}$ in M\o ller scattering is proportional to $Q_W^eG_FQ^2$~\cite{Derman:1979zc}
and is highly suppressed. Firstly, the electron's weak charge $Q_W^e\approx -1+4\sin^2\theta_W$ is very small. 
Further, while 
sufficient luminosity can be generated by utilizing a very dense target, $Q^2$ for high energy electrons 
scattering off electrons of mass $m_e$ 
in a fixed target is $\sim m_eE_{beam}$ and also very small. While $Q^2$ can be increased
by several orders of magnitude in collider mode, it is difficult to compensate for the larger loss in luminosity.

A feasible design concept was conceived~\cite{Kumar:1995ym}  after the upgrade of the SLAC linac 
enabled high intensity delivery of 48 GeV beam for the SLAC Linear Collider. That allowed the
first successful M\o ller $A_{PV}$ measurement by the E158 experiment~\cite{Anthony:2005pm}. 
However, even with a nearly 50 GeV beam,
the predicted value of $A_{PV}$ is only about 100 parts per billion (ppb) and many technical developments
were required in the production and monitoring of the highly polarized electron beam, in a high luminosity target, a
novel spectrometer and detection techniques. We now elaborate on the basic experimental 
technique for measuring $A_{PV}$ in fixed target polarized electron scattering, setting the stage for the description
of future initiatives using the same technique in Sec.~3.

The experiment was performed with 45 or 48 GeV
polarized electrons in 100 ns bunches at a rate of 120 Hz.  All
electrons scattered from a 1.5 m long hydrogen target with angles between 4.4 and 7.5 mrad
and energies between 13 and 24 GeV were focused onto a copper and quartz fiber calorimeter by a
quadrupole spectrometer.  The helicity of the beam was reversed from pulse to pulse in a pseudo-random pattern by using a Pockel's cell to reverse the helicity of the laser which produced the polarized electrons by photoemission from a strained GaAs crystal.  
The electron beam polarization was extracted via dedicated calibration runs measuring 
M{\o}ller scattering from a thin magnetized foil.

Care was taken to eliminate  false asymmetries due, for example, to differences in
beam properties correlated with helicity.  The position, angle and energy of the
beam were monitored with nanometer sensitivity and 
small corrections were made based on regular calibrations.  
In addition, the asymmetry was reversed every few runs by inserting a half-wave plate into the laser beam.  Finally, the helicity of the beam was opposite for each of the two energies due to a $g-2$ flip in the magnets in the beam switchyard; hence the data were collected with roughly equal statistics at the two different beam 
energies (45 and 48 GeV).  The asymmetry had the same magnitude and the correct sign for each of the four running configurations (half-wave plate state and beam energy), giving confidence
in the results and suppressing a wide variety of possible small spurious effects.

The result of the experiment was $A_{PV}=-131\pm14~{\rm (stat)}\pm10~{\rm (sys)}$ ppb. 
The tree-level prediction for $A_{PV}$ at the specific experimental
kinematics of E158 is about 250 ppb; the measured result demonstrated the running of $\sstw$ unambiguously
(more than 6$\sigma$) for the first time. 
The 3\%\ shift in the running of $\sstw$ to low $Q$ noted in Sec.~\ref{seclowqsin} 
results~\cite{Czarnecki:1995fw} in a shift in $Q_W^e = -1+4\sstw$ of about 40\%. 
Care was taken to include full electroweak radiative corrections, including
hard bremsstrahlung in the kinematic coverage~\cite{Kolomensky:2005ja}, yielding the value
of $\sin^2\theta_W$ quoted in Eqn.~\ref{eqthirtysix}, which stands currently as the best measurement
at $Q^2\ll M_Z^2$. To extract a value for $Q_W^e$, it is first necessary to define it unambiguously; see
Sec.~\ref{sectheory} for a full discussion. A logical choice, similar to the case of atomic parity violation discussed
earlier, is to define $Q_W^e$ in the static limit $E$ and $Q^2\to 0$. From the E158 result, the extracted
value is $Q_W^e = -0.0369(52)$.  In Sec.~\ref{secnew}, the resulting limits on four-electron contact interactions and its complementarity to similar limits from lepton colliders will be discussed.

\subsection{NuTeV}

The NuTeV experiment carried out the most precise measurement of neutrino neutral
current scattering utilizing neutrino beams of high energy and purity 
produced from the 800 GeV proton beam at Fermilab.
The weak mixing angle $\sin^2\theta_W$ was determined by measuring
the ratios of neutral to charged current cross sections in deep inelastic scattering for
both neutrinos ($R_\nu$) and anitneutrinos ($R_{\overline\nu}$)~\cite{Zeller:2001hh}. 
By using the ratios of cross sections, a
major experimental uncertainty, the details of the composition of the neutrino
beams, is largely canceled.  The events were detected in an 18 m long
steel-scintillator calorimeter followed by an iron-toroid spectrometer.
Because the target is approximately
isoscalar, the parton distribution functions (pdf's) largely cancel in the ratio, reducing theoretical
uncertainties.

One potential source of theoretical uncertainty for extracting $\sin^2\theta_W$
is the production of charm quarks via charged current interactions with strange sea quarks.  
The pdf $s(\xi)$, where $\xi=x(1+m_c^2/Q^2)$ is a slow rescaling variable, must be used for the charged current
whereas $s(x)$ is used for the neutral current. However, by treating both $R_\nu$ and $R_{\overline\nu}$ as
functions of $m_c$ and $\sin^2\theta_W$, and combining both measurements, the $m_c$ error can be reduced.
In the simplest approximation, a linear combination of $R_\nu$ and $R_{\overline\nu}$ 
can be found that is independent of $m_c$ and
is proportional to $1-2\sin^2\theta_W$, the Paschos-Wolfenstein relation~\cite{Paschos:1972kj}.

The published NuTeV result is nearly 3$\sigma$ away from the SM prediction, though some small shifts in either
direction are expected from updates to the $K_{e3}$ branching ratio, radiative corrections and isospin-breaking
effects. A number of phenomenological approaches
exploring physics beyond the SM have been investigated to interpret the discrepancy; for a review, see 
for example Ref.~\cite{Davidson:2001ji}.

A number of papers have been published trying to explain the NuTeV result
within the context of the SM.  One example invokes an asymmetric
quark sea, parton-level charge symmetry violation (CSV), and a modification of light quark pdf's in the nuclear
medium (so-called isovector EMC effect) to bring the experiment into
perfect agreement with the SM~\cite{Bentz:2009yy}. Other possibilities include radiative
corrections~\cite{Diener:2005me} and nuclear shadowing~\cite{Hirai:2004ba,Brodsky:2004qa}.  
Such corrections have not been incorporated formally into a reanalysis of the NuTeV result because of 
concerns about the estimated theoretical uncertainties of various corrections.  
If CSV and the isovector EMC
effect are indeed as large as given in Ref~\cite{Bentz:2009yy}, 
it would be a significant
discovery regarding fundamental QCD effects in nuclei.  
One of the auxiliary measurements in the proposed SoLID experiment, discussed in the
next section, would provide independent confirmation of this effect.

\section{PARITY-VIOLATING ELECTRON SCATTERING}
\label{secprogram}

In the following section, we describe the current experimental program of parity-violating electron scattering experiments. They are centered at two laboratories: the Thomas Jefferson National Accelerator Facility in 
Newport News, VA (JLab), and at the Institut f\"{u}r Kernphysik at the University of Mainz, Germany. 
All the experiments described here make
use of and build on 
the experimental techniques developed and improved over the decades since the pioneering SLAC E122 
experiment; an overview was provided in Sec.~2.3 in the description of SLAC E158.

The Continuous Electron Accelerator Facility (CEBAF) at JLab has been operating since 1995 with a wide dynamic range in beam energy (from 1 to 6 GeV), beam current (few nA to 180 $\mu$A),
longitudinal beam polarization ($> 85$\%) and beam stability. In 2014, an energy upgrade will be completed which
will increase the maximum available beam energy to 12 GeV, with the capability of delivering 11 GeV at very 
high luminosity to existing experimental halls, significantly expanding the physics 
program~\cite{Dudek:2012vr}. 
The Qweak experiment recently completed data collection using
a 1 GeV beam energy and two new initiatives known as MOLLER and SoLID have been proposed to utilize the 11 GeV beam.

The Mainz Energy-recovering Superconducting Accelerator (MESA) is a new machine which has been approved
for funding at Mainz, offering 100 MeV in energy recovery operation and 150-200 MeV for conventional 
external beam mode~\cite{Aulenbacher:2012tg}. 
The latter mode is the one that will be used for the proposed P2 experiment. 
It is envisioned that first beam will be available by the end of 2017. 

\subsection{Qweak}
\label{secQweak}

The Qweak experiment~\cite{Armstrong:2012ps} 
was designed to measure the proton's weak charge $Q_W^p \approx 1-4\sstw$ via $A_{PV}$ in elastic electron-proton scattering. The experiment was first proposed in 
2001, constructed between 2006 and 2009, and data collection was completed in 2012 in two run periods
lasting about 11 months in total. 
Data analysis is ongoing and final results are expected by 2014.

The experimental design centered around achieving $\delta(A_{PV})\approx \pm 2.1$\%\ (stat.) and 
$\pm 1.3$\% (syst.), resulting in $\delta(Q_W^p)\approx\pm 4$\%, and 
$\delta(\sin^2\theta_W)\approx \pm 0.3$\%. 
The incident beam energy was 1.165 GeV. Elastically scattered electrons in the range 
$\theta_{lab} = 8\pm 2^\circ\rightarrow\langle Q^2\rangle = 0.026$ GeV$^2$ were selected. 
The theoretical prediction at this $Q^2$ is $A_{PV}\approx -230$ ppb, the piece proportional to $Q_W^pG_F$
is $-150$ ppb and the statistical goal was $\delta(A_{PV})\approx \pm 6$ ppb. 

%The experimental layout is shown Fig.~\ref{}. 
In Hall C at JLab, a 1 GeV 87\%\ longitudinally polarized electron beam, with a 
current of $150-180 \mu$A, was incident on a 35 cm liquid hydrogen target capable of withstanding a heat load
of 2.5 kW. Elastically scattered electrons were focused by the spectrometer/collimator system on to an 
azimuthally symmetric (with respect to the beam axis) arrangement of quartz 
bar integrating Cherenkov detectors. The electron beam helicity was reversed in a quartet pattern at 960 Hz:
a helicity state was chosen pseudo-randomly at 240 Hz and 4 consecutive pulses in  $(+--+)$ pattern or its
complement were chosen accordingly. The width of the raw asymmetry distribution is a crucial bench mark; 
for Qweak, this width was 230 ppm for a quartet. 
The contribution from counting statistics was about 200 ppm. The dominant sources of additional fluctuations were from detector resolution, target density fluctuations and beam current 
monitor resolution. 

Methods similar to those developed for SLAC E158 (Sec.~2.3) were employed to reduce the sensitivity 
of the measured asymmetry to helicity-correlated beam fluctuations. Also, a new method for reversing the
relative direction between the spin and momentum vectors of the incident electrons before acceleration (so-called
Double Wien filter) was
employed every few weeks to gain further suppression. In addition to using M\o ller polarimetry every few days, 
Qweak used a Compton polarimeter that monitored the electron beam polarization continuously, concomitant 
with physics data collection; it is anticipated that the absolute beam polarization will be known to better than 1\%. The absolute value of $\langle Q^2\rangle$ was calibrated in separate low current runs using special purpose drift-chambers that could track individual scattered electrons. 

\subsection{MOLLER}
\label{secMOLLER}
The MOLLER experiment~\cite{Mammei:2012ph} 
is a new initiative proposed to measure $A_{PV}$ in M\o ller scattering
a factor of 5 better than the E158 result. As was pointed out in Sec.~\ref{secE158}, 
$A_{PV}\propto Q_W^e\approx -1+4\sstw$ which is reduced from its tree-level value 
by $\sim 40\%$ due to radiative corrections. This reduces the
sensitivity of the extracted value of $\sstw$ to normalization errors such as beam polarization 
by an additional factor of two compared to $A_{PV}$ in elastic electron-proton scattering.
The goal is a 2.3\%\ measurement of $Q_W^e$ resulting in $\delta(\sin^2\theta_W)_{\rm stat}\approx 0.00025$, 
${\cal{O}}(\pm0.1\%)$, similar to the two best high energy collider 
extractions of the parameter from measurements of $\zzero$ decays (see Sec.~1).

The MOLLER design shares many similarities with E158 and Qweak. 
%A schematic diagram of the experimental layout  is shown in Fig.~\ref{}. 
An 11 GeV electron beam in JLab Hall A will be incident on a 1.5
meter LH$_2$ target. A toroidal spectrometer would exploit 
the unique topology of M\o ller scattering involving identical particles, avoiding
the typical 50\%\ azimuthal acceptance loss associated with coil placement. 
This is accomplished by employing an odd number of coils and collecting scattered electrons 
from both the forward and backward directions in the center of mass frame. 
The M\o ller-scattered electrons in the full range of the azimuth would be directed to a ring focus 30 m 
downstream of the target. The detector system would incorporate a great deal of redundancy to monitor the
principal backgrounds from electron-proton elastic and inelastic scattering to better than 1\% accuracy. 
The prediction for $A_{PV}$ is 35.6 ppb and the statistical error goal is 0.74 ppb. 

MOLLER will greatly benefit from the steady improvement in the techniques employed to measure parity-violating 
asymmetries to sub-ppb systematic precision and to also achieve normalization control at the sub-\%\ level. For
example, two redundant continuous monitors of electron beam polarization would be employed  to achieve
0.4\% fractional accuracy. Auxiliary detectors would track individual particles at low rates to measure $\langle
Q^2\rangle$ to 0.5\% fractional accuracy. Very forward angle detectors downstream of the main detectors would 
verify that 
luminosity fluctuations due to jitter in electron beam properties and target density are under control. 
All three methods
to reverse the sign of the asymmetry that have been developed for previous experiments ($g-2$ spin flip, half-wave
plate insertion and the double Wien filter) would also be employed periodically.
Technical design efforts for MOLLER are ongoing and it is envisioned that funding approval will be obtained
in 2013 so that construction of the apparatus could begin in 2015 which would allow commissioning by 2017, 
soon after full luminosity beams become available at Jefferson Laboratory.

\subsection{Deep Inelastic Scattering at 6 GeV}

The first measurement of $A_{PV}$ in deep-inelastic scattering since the original SLAC E122 measurement 
discussed in Sec.~\ref{sec:history}~\cite{prescott78} was carried out by JLab experiment 
E08011~\cite{Zheng:2012vf}. 
The primary motivation was to measure the poorly known neutral current axial-vector quark
couplings. 
The experiment ran in late 2009 with an incident beam energy of 
$\sim 6$ GeV and $Q^2\sim 1-2$ GeV$^2$, collecting sufficient statistics to measure $A_{PV}$ off $^2$H with a 
fractional accuracy better than 4\%. 

The scattered electrons were detected by the Hall A High Resolution Spectrometer (HRS) 
pair~\cite{Alcorn:2004sb}. Unlike other high rate experiments discussed in this review, a custom fast counting
data acquisition system was used. Event-by-event particle identification was carried out at the hardware level 
with gas Cherenkov detectors and lead-glass shower counters. This information was fed into fast trigger logic
to form electron and pion triggers that were in turn fed into scalers. The electron scaler results over the duration
of each helicity time window of the electron beam were used to construct the raw asymmetry from which $A_{PV}$
could be extracted. The electron trigger efficiency was found to be greater than 95\%, with a pion rejection 
$> 1000:1$. Data analysis is ongoing and final results are expected to be published by late 2013.

\subsection{SOLID}
\label{secSOLID}

The SoLID experiment~\cite{Souder:2012zz} at JLab has been proposed to make a series of $A_{PV}$
measurements with 0.5--1\%\ fractional accuracy in deep inelastic scattering of electrons off $^2$H. 
The primary motivation is to measure new linear combinations of vector and axial-vector quark couplings
with sufficient accuracy to provide new and complementary access to new TeV-scale lepton-quark interactions. 
It would
also result in a measurement of $\sstw$ with an uncertainty of $\delta(\sstw)\approx 0.0006$ at $Q\sim 2.5$ GeV.

The heart of the apparatus is a large acceptance solenoid such as one of several that have been 
used over the past couple of decades to provide the magnetic field inside collider detectors. This 
facilitates $A_{PV}$ measurements in narrow $x_{\rm bj}$, $Q^2$ bins:
$0.3 \lesssim x_{\rm bj} \lesssim 0.7$, with a lever arm of a factor of 2 in 
$Q^2$ while keeping $W^2_\mathrm{min}>4$ GeV$^2$ and $\langle Q^2\rangle\sim 5$ GeV$^2$. 
Such a large volume and high field solenoid is required to achieve sufficient statistics at the highest 
possible $Q^2$ and $x_{\rm bj}$,  provided 
at least 50\%\ azimuthal acceptance is obtained. It facilitates shielding the detectors from
target photons and sweeping out low energy charged particles, while accommodating a significant target length
and large laboratory scattering angles. 
The $\mathrm{LD}_2$ target would be placed inside
the solenoidal field and several planes of absorbing material between the target and detectors with slits would 
tailor the momentum acceptance to the kinematic region interest.  
%A diagram of the apparatus is shown in Fig.~\ref{fig:solidapp}.

One important feature is that, unlike other $A_{PV}$ measurements that integrate detector signals over
different helicity periods, high precision hit-based tracking with gas electron multiplier detectors is 
required to reconstruct the scattering angle and momentum of scattered electrons.  
To separate electrons from background, predominantly a 100 times more $\pi^-$ particles, particle 
identification would be performed with heavy-gas 
C$\hat{e}$renkov detectors placed symmetrically about the beam axis inside the solenoid. An 
electromagnetic calorimeter would provide the primary electron trigger as well as allow additional pion rejection. 

A proposal for the experiment 
was approved in January 2010 at JLab and detailed simulations
have been carried out for the case of using the solenoid from the CLEO-II 
detector in the CESR e$^+$e$^-$ storage ring.
A significant R\&D program has been launched to develop a detailed 
experimental design and the project will seek funding over the next few years to run at JLab.

\subsection{P2}
\label{sec:P2}

The P2 experiment has been proposed for the newly funded MESA facility at Mainz. The goal
is $\delta(A_{PV})= \pm 1.7\%$ (stat. + syst.) for elastic electron-proton, which would yield 
$\delta(Q_W^p)\simeq 2\%$ and $\delta(\sin^2\theta_W)\pm 0.15\%$. 
To achieve the statistics would require
a 200 MeV, 150 $\mu$A beam incident on a 60 cm LH$_2$ target for 10,000 hours.
Apart from the improvement
in statistical reach and hence sensitivity to new physics over the JLab Qweak experiment, the lower 
beam energy reduces theoretical uncertainties in extracting $\sin^2\theta_W$ (see Sec.~\ref{secqweaktheory}). 

The design requires a solenoidal magnet (such as the inner tracking field of the ZEUS collider detector at 
DESY) downstream of the target which would focus scattered electrons within 
$10^\circ<\theta_{\rm lab}<30^\circ$ in the full range of the azimuth onto integrating Cherenkov
detectors.  The field would sweep out the large M\o ller electron background and allow judiciously placed 
annular slits to shield the detectors from the target's direct photon background. 
The theoretical prediction is $A_{PV}\sim 20$ ppb and $\delta(A_{PV})$(stat.) is $\pm 0.25$ ppb. The total rate in the detectors would approach 0.5 THz. 

The design must overcome many technical challenges such as controlling electron beam fluctuations at the
sub-nm level and controlling target density
fluctuations to a few parts in $10^{-5}$. A new method to measure the electron beam 
polarization must be developed, which would require a novel polarized hydrogen 
gas target~\cite{Chudakov:2004de}.
The design and required R\&D will be carried out in the next few years so that the experiment
would be ready to start commissioning when MESA first produces external beams, anticipated for 2017.

\section{SENSITIVITY TO PHYSICS BEYOND THE STANDARD MODEL}
\label{secnew}

We now discuss the sensitivity of precision low $Q^2$ measurements of weak neutral current 
amplitudes to physics beyond the SM. We choose a few specific topics; many comprehensive 
reviews~\cite{RamseyMusolf:1999qk, beringer12} have studied aspects of the 
sensitivity to supersymmetric~\cite{RamseyMusolf:2006vr, erlersu} as well as 
non-supersymmetric~\cite{Chang:2009yw} new physics. 

\subsection{New Contact Interactions}
If there is new physics beyond the SM at some scale $\Lambda$ above the electroweak scale, then in  
measurements at $Q^2\ll\Lambda^2$ new dynamics can manifest themselves as small deviations 
from the expected SM rates.  The new dynamics appear as contact interaction terms in an effective 
Lagrangian~\cite{Eichten:1983hw} that interfere with SM amplitudes. They can be parametrized as:
\bea
\label{highdim}
{\cal L}_{\mbox{eff}} &=& \frac{g^2}{(1+\delta) \Lambda^2} \sum_{i,j=L,R} \> \eta_{ij}^f \bar{e}_i\gamma_\mu e_i \bar{f}_j\gamma^\mu f_j,
\eea
summed over helicities ($\delta=0(1)$ for $f=e (f\neq e)$). A typical convention sets $g^2/(4\pi)=1$
and the $\eta_{ij}^f=\pm1$ or $0$. Precision measurements can then be translated into bounds on $\Lambda$.
The effective Lagrangian in Eq.(\ref{highdim}) can be induced by a range of new physics scenarios such as low
scale quantum gravity with large extra dimensions, composite fermions, leptoquarks, heavy $Z'$ bosons etc.

One can classify models that induce contact interactions according to the choices for the $\eta^f_{ij}$;
a representative sample~\cite{Kroha:1991mn} is shown in Table \ref{Leffmod}.
\begin{table}
\begin{minipage}[b]{0.45\linewidth}
\begin{center}
  \begin{tabular}{ | l | c |c | c | c | c | r | } 
    \hline
    \mbox{Model} & $\eta^f_{LL}$ & $\eta^f_{RR}$ &$\eta^f_{LR}$ &$\eta^f_{RL}$   \\ \hline \hline
    $LL^{\pm}$ & $\pm1$ & 0 & 0 & 0 \\ \hline
    $RR^{\pm}$ & 0 & $\pm1$ & 0 & 0 \\ \hline
           $LR^{\pm}$ & 0 & 0 & $\pm1$ & 0 \\ \hline
        $RL^{\pm}$ & 0 & 0 & 0 & $\pm1$ \\ \hline
     $VV^{\pm}$ & $\pm1$ & $\pm1$ & $\pm1$ & $\pm1$ \\ \hline
      $AA^{\pm}$ & $\pm1$ & $\pm 1$ & $\mp 1$ & $\mp 1$ \\ \hline
                $VA^{\pm}$ &   $\pm1$ & $\mp1$ & $\pm1$ & $\mp1$ \\ \hline
%         $V0^{\pm}$ & $\pm1$ & $\pm1$ & 0 & 0 \\ \hline
%          $A0^{\pm}$ &   0 & 0 & $\pm1$ & $\pm1$ \\ \hline
  \end{tabular} 
\end{center}
\caption{Models classified by chiral structure in the effective Lagrangian.}
\label{Leffmod}
\end{minipage}
\hspace{0.2cm}
\begin{minipage}[b]{0.49\linewidth}
\begin{center}
  \begin{tabular}{ | l | c | l | } 
    \hline
    \mbox{Experiment} & $\Lambda$  & Coupling \\ \hline \hline
     Cesium APV  & 9.9 TeV & $C_{1u} + C_{1d}$ \\ \hline
    E-158 & 8.5 TeV & $C_{ee}$ \\ \hline
           Qweak & 11 TeV & $2C_{1u} + C_{1d}$ \\ \hline
           SoLID & 8.9 TeV & $2C_{2u} - C_{2d}$\\ \hline
    MOLLER & 19 TeV & $C_{ee}$ \\ \hline
        P2  & 16 TeV & $2C_{1u} + C_{1d}$\\ \hline
  \end{tabular} 
\end{center}
\caption{95\%\ C.L. reach of experiments discussed in Sec.~\ref{secpast} and~\ref{secprogram} 
to the new physics scale $\Lambda$ ($g^2 = 4\pi$)}
\label{NPlimits}
\end{minipage}
\end{table}
The superscripts $(+,-)$ indicate constructive and destructive interference with the
SM respectively.  Searches for such new contact interactions have been carried out in electron-positron, 
electron-proton, and hadron colliders. 

The $LL^\pm$ model, shown in Table \ref{Leffmod}, is a benchmark scenario commonly used in contact interaction searches. For the $eeqq$-type contact interactions, assuming quark flavor independence, sensitive limits come 
from analyses of cross-sections and asymmetries at LEP~\cite{Schael:2006wu, Schael:2013ita}: 
$\Lambda^- >$  8.0 TeV and $\Lambda^+ >$  9.7 TeV ($95\%$ C.L.). Similarly, 
$\Lambda^- >$  7.0 TeV and $\Lambda^+ >$  4.5 TeV were obtained for $eeee$-type contact interactions.
More comprehensive bounds for several of the other models in Table \ref{Leffmod} can be found in 
Ref.~\cite{ALEPH:2004aa}. More recently, the ATLAS \cite{:2012cb} collaboration  derived bounds on the 
$LL^\pm$ model via measurements of fully-inclusive Drell-Yan production in the dielectron channel, obtaining
$\Lambda^- > $ 9.5 TeV, $\Lambda^+ >$ 12.1 TeV for the  $LL^\pm$  $eeqq$-type contact interaction. 

Low $Q^2$ weak neutral current measurements discussed in 
Secs.~\ref{secpast} and~\ref{secprogram} can also probe contact interactions competitive with colliders.
Specifically, the parity violation measurements are sensitive to contact interactions of the form
\bea
\label{Leff}
{\cal L } &=& -\frac{G_F}{\sqrt{2}} \sum_{q} \Big [C_{1q}\>\bar{e} \gamma^\mu\gamma_5 e\>   \bar{q}\gamma_\mu q + C_{2q}\>\bar{e} \gamma^\mu e\>   \bar{q}\gamma_\mu \gamma_5 q\Big ] \nn \\
&-& \frac{G_F}{\sqrt{2}}  C_{2e} \>\bar{e} \gamma^\mu \gamma_5 e\> \bar{e} \gamma_\mu  e,
\eea
where the sum over $``q"$ is over the quark flavors and the coefficients $C_{1q}, C_{2q}, C_{2e}$ are given by the sum of SM and new contact interaction contributions. Bounds on specific couplings can be 
translated into bounds on $\Lambda$. For example, the E158 measurement (Sec.~\ref{secE158})
can be used to extract a value of the weak charge of the electron (in the static limit, see Sec.~\ref{sectheory})
$Q_W^e = 2C_{2e} \approx -1+4\sstw = -0.0369(52)$ compared to the SM prediction of -0.0435(9). One 
can then extract 95\%\ C.L. limits on $\Lambda$ using
\bea
\label{Qshift}
\Lambda \simeq  \frac{2\sqrt{\pi}}{\sqrt{\sqrt{2} G_F \Delta Q_W^e}}
\eea
to obtain $\Lambda_{LL}^+\geq 6.7$ TeV and $\Lambda_{LL}^-\geq 14.2$ TeV. Two things are worth pointing
out. Firstly, the sensitivity is better than the limits from LEP, underscoring the power
of measuring a small SM coupling such as $C_{2e}G_F$ to high precision. Secondly, the LEP limits come
from measurements above $W^+W^-$ threshold. The precise $\sstw$ measurements at the $\zzero$
resonance are not as sensitive to contact interactions amplitudes; the imaginary SM amplitude on top of the
$\zzero$ resonance does not interfere with them. 
The proposed MOLLER measurement (Sec.~\ref{secMOLLER}) would
improve these $\Lambda$ limits to nearly 20 TeV, the best sensitivity to new flavor-conserving four-lepton
contact interactions  in existing facilities anywhere in the world. Better limits would require
the construction of new facilities such as a linear collider, $\zzero$ factory or neutrino factories, all of which
are at least a decade away from fruition.

The chiral structures for $eeqq$-type contact interactions that appear in Eqn.~\ref{Leff} correspond 
to the $AV^\pm$ ($C_{1i}$) and $VA^\pm$ ($C_{2i}$)
class of models in Table \ref{Leffmod}. Limits on the $VA^\pm$ class of $eeqq$-type 
contact interactions in Table \ref{Leffmod} were
 obtained by the H1~\cite{Aaron:2011mv} and Zeus~\cite{Chekanov:2003pw} experiments studying  
 electron-proton and positron-proton collisions in the deep inelastic regime to give  
 $\Lambda^- > $ 3.6 TeV, $\Lambda^+ >$ 3.8 TeV and  $\Lambda^- > $ 3.2 TeV, $\Lambda^+ >$ 3.3 TeV
respectively.  

$A_{PV}$ and APV measurements in semi-leptonic reactions discussed in 
Sec.~\ref{secpast} and Sec.~\ref{secprogram} have complementary and improved sensitivity. Over the last two decades, several experiments have measured $A_{PV}$ in elastic electron-proton scattering
at $0.1<Q^2<1$ GeV$^2$ with the aim of constraining the strange quark form factors of the 
proton~\cite{Kumar:2000eq}. Global fits
to the data determined that the strange form factors were constrained to be small 
enough~\cite{Young:2006jc, Liu:2007yi, GonzalezJimenez:2011fq} 
that the low $Q^2$ forward angle data could then be analyzed to extract
a measurement of $Q_W^p=2(2C_{1u}+C_{1d})$. 
By expanding the parity-violating asymmetry at small scattering angles in powers of $Q^2$ with the parametrization
$A_{PV}\propto Q_W^p + Q^2 B(Q^2)$,
and then combining with the atomic parity violation result on Cesium discussed in Sec.~\ref{secAPV} (which
measures $C_{1u} + C_{1d}$), independent 
determinations of $C_{1u}$ and $C_{1d}$ were obtained~\cite{Young:2007zs}. This led to new constraints on 
the $AV^\pm$ chiral structure of $\sim 3$ TeV
independently for $eeuu$- and $eedd$-type interactions, comparable to the H1 and ZEUS limits.

The Qweak measurement (Sec.~\ref{secQweak}) will improve the sensitivity to the specific linear combination
of $AV^\pm$ interaction $2C_{1u}+C_{1d}$ to better than 10 TeV, while the APV result on Cesium already has
similar sensitivity reach for $C_{1u}+C_{1d}$. The SoLID measurement would have sensitivity to a 
new linear combination of $VA^\pm$ $eeqq$-type contact interactions at the level of 8.9 TeV. Note that
improving $\Lambda$ sensitivity beyond 10 TeV in a variety of chiral structures is 
necessary for a comprehensive search, as demonstrated by an example discussed at the end of the next section.
The $\Lambda$ reach and the specific coupling combinations of various experiments  discussed in this review are 
summarized in Table~\ref{NPlimits}.

\subsection{New Heavy $Z'$ Bosons}
$Z'$ bosons with mass $M_Z'$ in the TeV range appear in many extensions of the SM, including $SO(10)$, $E_6$, 
Little Higgs, and extra-dimensional theories.  They arise from an additional $U(1)'$ gauge group appearing in such
new physics  constructions. The phenomenology has been extensively 
reviewed~\cite{Rizzo:2006nw,Langacker:2008yv}, and the impact of precision electroweak data on a 
wide range of $Z'$ models extensively analyzed~\cite{Erler:2009jh,Erler:2011ud}. 
$Z'$s have been constrained by electroweak precision data~\cite{Erler:2009jh, delAguila:2010mx}, interference
effects at LEP-II~\cite{ALEPH:2004aa}, and the Tevatron~\cite{Jaffre:2009dg} with limits $\sim 1$ TeV.
 
The simplest discovery mode for $Z'$ bosons would be through an $s$-channel resonance in the dilepton or dijet 
topologies at colliders.  The LHC will be able to explore the 1--5 TeV range of $Z'$ masses, although extracting 
detailed properties such as the width and couplings  will be difficult for $M_Z' \gtrsim 2$ TeV. In the region
$M_Z'\lesssim 2$ TeV,  a detailed study of the couplings will be enhanced by an analysis of off-peak LHC data and
low-energy electroweak precision data. In such analyses, where the interaction energies are well below  the $Z'$
mass ($M_{Z'}$), its exchange can be described by contact interactions where $\Lambda \sim M_{Z'}$. Constraints 
on contact interactions can then be translated into bounds on the mass and couplings of the $Z'$ boson.
For example, for $Z'$ exchange between electrons and quarks, with the $LL^\pm$ chiral structure of 
Table~\ref{Leffmod},  $M_{Z'}^2/(g_e^L g_{q}^L) \simeq \Lambda^2/(4\pi^2)$ where $g_{e}^L,g_{q}^L$ are the left-handed $Z'$ couplings to electrons and quarks respectively.

Low $Q^2$ $A_{PV}$ measurements can be quite sensitive to $Z'$ bosons due to their sensitivity to the interference
between the electromagnetic amplitude and the $Z'$ contact interaction. Further, they probe different
chiral combinations of $Z'$ couplings compared to the LHC, helping remove degeneracies in parameter space in 
a purely LHC data-based analysis~\cite{Li:2009xh}. In the context of exploring the reach of low energy experiments,
a class of $E_6$ based $Z'$-models was recently  analyzed~\cite{Erler:2011iw}. Such $Z'$ models arise from the spontaneous symmetry breaking chain $E_6 \to SO(10) \times U(1)_{\psi} \to SU(5) \times U(1)_\chi \times U(1)_{\psi}\to \mbox{SM} \times U(1)'$. The $Z'$ associated with the remaining $U(1)'$ can be written in general form
\bea
\label{zprime}
Z'=\cos \alpha \cos \beta Z_\chi + \sin \alpha \cos \beta Z_Y + \sin \beta Z_\psi,
\eea
where $Z_{\chi,Y,\psi}$ are gauge bosons associated with $U(1)_{\chi,Y,\psi}$ in the gauge eigenstate basis, 
$\beta$ is the mixing angle between $U(1)_{\chi}$ and $U(1)_{\psi}$, and 
the angle $\alpha$ is non-vanishing in the presence of kinetic mixing between the $U(1)'$ and $U(1)_Y$ gauge 
groups. The angles $\alpha$ and $\beta$ in Eq.(\ref{zprime}) provide a way to parameterize the class of $E_6$
based models. For example, the $Z_\chi, Z_\psi,$ and $Z_\eta$ models correspond to $\alpha=0$ and 
$\beta=0^0, 90^0,-\mbox{arctan}\sqrt{5/3}$ respectively. An example of how low energy results and future initiatives
discussed in this review complement collider searches is demonstrated in Fig.~1 of Ref.~\cite{Erler:2011iw}, which
shows excluded regions in the $(\alpha,\beta)$ parameter space for a hypothetical $M_Z'=1.2$ TeV.

One particularly unique sensitivity of low $Q^2$ $A_{PV}$ measurements is with that to the so-called lepto-phobic
$Z'$ boson that only couples to quarks and is thus difficult to discover at hadron colliders due to the irreducible QCD 
backgrounds. As mentioned in the previous section, measurements of the $C_{2i}$ couplings by the proposed
SOLID experiment (Sec.~\ref{secSOLID}) will provide new sensitivity to $VA^\pm$ $eeqq$-type contact interactions.
Recently, it has been pointed out~\cite{Buckley:2012tc, GonzalezAlonso:2012jb} that these couplings can be modified by a $\gamma-Z'$ vacuum-polarization one-loop correction, thus extending the current $< 100$ GeV reach to $150-200$ GeV.

\subsection{Dark Parity Violation}
\label{secdarkz}

Some recent new physics scenarios have relatively light new degrees of freedom and hence cannot be classified in 
terms of contact interactions. The failure to observe such scenarios in high energy experiments implies
that such light particles must couple very weakly to SM particles.  In certain regions of parameter space, low energy precision experiments can have unique or enhanced sensitivity. In this context,  the possibility of a ``dark" $Z$ boson~\cite{Davoudiasl:2012ag,Davoudiasl:2012qa}, denoted as 
$Z_d$ and of mass $m_{Z_d}$,  stemming from a spontaneously broken $U(1)_d$ gauge symmetry associated with
a secluded ``dark" particle sector was recently investigated. The $Z_d$ boson can couple to the SM through
a combination of kinetic and mass mixing with photon and 
the $Z^0$-boson, with couplings $\varepsilon$ and $\varepsilon_Z = \frac{m_{Z_d}}{m_Z}\delta$ respectively. 

The original scenario with kinetic mixing with the photon was 
conjectured~\cite{Holdom:1985ag, Fayet:2004bw, Bouchiat:2004sp, Pospelov:2008zw} to 
explain astrophysical observables as well as to account for the long standing deviation of the muon's anomalous magnetic moment, $a_\mu$, from SM expectations~\cite{Bennett:2006fi}.
Taking into account various experimental constraints, the $a_\mu$ discrepancy is naturally accommodated by kinetic mixing in the range
\beq
|\varepsilon| \simeq 2\times 10^{-3} \qquad\qquad 20{\rm~MeV}\lsim m_{Z_d}\lsim 50 {\rm~MeV} \label{eqdarktwo}.
\eeq

In the presence of mass mixing ($\delta \neq 0$),  a new source of ``dark'' parity violation arises~\cite{Davoudiasl:2012ag} such that it has negligible effect at the $\zzero$ pole precision data, 
but is quite discernable at low $Q^2$ through a shift in the weak mixing angle~\cite{Davoudiasl:2012qa}:
\beq
\Delta\sstw(Q^2)\simeq-0.42 \varepsilon\delta \frac{m_Z}{m_{Z_d}} \left( \frac{m^2_{Z_d}}{Q^2+m^2_{Z_d}}\right). \label{eqdarkthree}
\eeq
\noi In this scenario, the small ($1.5\sigma$) APV deviation, 
$\Delta\sstw(0)\simeq -0.003(2)$ suggests $\delta\simeq\pm1-4\times10^{-3}$
as a potentially interesting region that can be explored by future APV or $A_{PV}$ measurements,
as we elaborate in the next section. 

\subsection{Weak Charges and New Physics}
\label{secnewweak}

Employing the very precisely measured values of $\alpha, G_F$, and $m_Z$ along with $m_t$ and $m_H=126$ GeV in the one-loop corrected SM, but allowing for very heavy new particle loop effects via the electroweak precision $S$ and $T$ parameters \cite{Peskin:1990zt,marciano90} leads to the predictions
\bea
m_W = 80.362(6) \mbox{GeV} \big [ 1-0.0036 \>S+0.0056 \>T \big ], \nn \\ 
\sin^2 \theta_W(m_Z)_{\bar{MS}} = 0.23124(6) \big [ 1+ 0.0157 S - 0.0112 \>T\big ].
\eea
Comparison with experiment currently yields $S=0.07 \pm 0.09$ and $T=0.10\pm 0.09$ which can be used to significantly constrain models such as Technicolor or the properties of 4th generation fermions.

Similarly, the weak charges of particles and nuclei (defined at the static limit, $E$ and $Q^2\to 0$) are precisely predicted at the loop level~\cite{Marciano:1983wwa}. 
However, in addition to $S$ and $T$, deviations can be induced by  new physics in other ways. For example, $Z'$ 
gauge bosons can cause ${\cal O} (m_Z^2/m_{Z'}^2)$ shifts in the weak charges similar to that seen in 
Eqn.~(\ref{Qshift}) for generic contact interactions. 

To illustrate relative sensitivities, we consider the $Z_\chi$ model 
of $SO(10)$ that violates parity in a well-specified manner~\cite{marciano90,London:1986dk}. 
Also, to allow new physics differences between 
$\sin^2\theta_W(m_Z)_{\mMS}$ and $\sin^2\theta_W(0)_{\mMS}$ beyond SM running effects, we introduce 
$X(Q^2)$~\cite{Maksymyk:1993zm,Czarnecki:1995fw}, which is similar to $S$, but $Q^2$-dependent. 
For heavy particle loops (eg. SUSY or heavy fermions~\cite{Kurylov:2003zh}) with a generic mass scale 
$M$ where $X(Q^2) \sim {\cal O}(\alpha Q^2/M^2)$, $X$ is already well constrained
by the bounds on $S$ from $W$ and $Z$ measurements. However, for very light new physics like the MeV-scale 
``dark" boson (Sec.~\ref{secdarkz}), $Z_d$, that mixes with $\gamma$ and $Z^0$, thus providing a new source
of dark parity violation~\cite{Davoudiasl:2012ag,Davoudiasl:2012qa},  
$X(Q^2)\propto m_{Z_d}^2/(Q^2 + m_{Z_d}^2)$  terms can occur that are 
only visible in low $Q^2\lesssim m_{Z_d}^2$ experiments.

Given these new physics scenarios, one finds the following shifts in the weak 
charges~\cite{marciano90, Czarnecki:1995fw, Blunden:2012ty}
\bea
\label{eqn:newweak}
Q_W^e &=& -0.0435(9) [1+ 0.25\>T-0.34\>S+0.7\>X(Q^2)+7m_Z^2/m_{Z_\chi}^2], \nn \\
Q_W^p&=& 0.0707(9)[1+0.15\>T-0.21 \>S+ 0.43\>X(Q^2) + 4.3 m_Z^2/m_{Z_\chi}^2], \nn\\
Q_W(^{12}C)&=& -5.510(5) [1-0.003\>T+0.016\>S-0.033\>X(Q^2)-m_Z^2/m_{Z_\chi}^2], \nn\\
Q_W(^{133}Cs)&=& -73.24(5) [1+0.011\>S-0.023\>X(Q^2)- 0.9 m_Z^2/m_{Z_\chi}^2],
\eea
\noi where the uncertainties have been somewhat expanded to account for as yet uncalculated 
higher order effects.
Several interesting features are apparent. The large $\sim 40\%$ radiative corrections to $Q_W^e$ improve its 
fractional sensitivity to new physics relative to $Q_W^p$. That makes M\o ller scattering better from a systematic
(such as polarization) perspective, but statistically similar in difficulty: $\delta(Q_W^e)\sim \pm 2\%$
is roughly equivalent to $\delta(Q_W^p)\sim \pm 1\%$. Both cases probe $S$ and $T$ below 
$\pm 0.1$ and $m_{Z_\chi} \sim 2$ TeV.

For nuclei such as $^{12}$C or $^{133}$Cs, the $T$-dependence is small \cite{marciano90}. 
Assuming $|S|\lesssim 0.1$ (based on existing constraints) suggests that they should be viewed as having mainly
$Z'$ and $X(Q^2)$ sensitivity. In the case of $Z_\chi$, a $\pm 0.3\%$ measurement of $Q_W(^{12}C)$ or 
$Q_W(^{133}Cs)$ is roughly equivalent to $\delta(Q_W^e)\sim \pm 2\%$. 

For $X(Q^2)$ effects such as due to low mass $Z_d$ particles \cite{Davoudiasl:2012ag,Davoudiasl:2012qa, Bouchiat:2004sp, Davoudiasl:2012ig} with $m_{Z_d}\sim 20 - 50$
MeV discussed in the previous section, APV experiments where $Q^2$ is naturally small ($Q^2 \ll m_{Z_d}^2$) are
 superior probes because they do not have the $m_{Z_d}^2/(Q^2 + m_{Z_d}^2)$ suppression. However, $Q_W^e$
and $Q_W^p$ are fractionally far more sensitive to $X(Q^2)$. For example, the $-0.9\pm 0.6\%$ shift in the APV Cs
result would lead to a 27\%\ shift in $Q_W^e$ if measured at the same $Q$. For $m_{Z_d}\sim 50$ MeV, the 
proposed MOLLER measurement ($Q\sim 75$ MeV) would see an 8.4\%\ shift ($\sim 3.7\sigma$). 
We later show that polarized eC scattering may also be a good probe of parity violating $Z_d$ effects if low 
$\langle Q\rangle$ can be achieved.

\section{SELECTED THEORETICAL ISSUES}
\label{sectheory}

\subsection{Radiative Corrections to Parity-Violating M\o ller Scattering}
\label{mollerrunning}

The asymmetry $A_{PV}$ in the M\o ller scattering process 
$e^-e^-\to e^-e^-$~\cite{Derman:1979zc} is a powerful probe 
of new physics with relatively small theoretical uncertainties. 
In the mid-1990's, the proposed precision of the E158 measurement (Sec.~\ref{secE158}) spurred the calculation
of one-loop corrections, which shifts $A_{PV}$ for M\o ller scattering by about 
40\%~\cite{Czarnecki:1995fw}. Indeed, given the plans to further improve on the E158 measurement 
(the MOLLER proposal, Sec.~\ref{secMOLLER}), 
significant progress has been made to control uncertainties in $A_{PV}$ from higher-order
radiative corrections at better than the 
1\%\ 
level~\cite{Czarnecki:1995fw,Denner:1998um,Petriello:2002wk,Erler:2004in,Aleksejevs:2010ub,Aleksejevs:2010nf}.
For $Q^2\ll m_Z^2$, the tree level expression is modified by radiative corrections as follows~\cite{Czarnecki:1995fw}:
\bea
\label{moller-radiative}
A_{PV} &=& -\frac{\rho G_F Q^2}{\sqrt{2}\pi \alpha} \frac{1-y}{1+y^4 + (1-y)^4} \big \{1-4\kappa(0) \sin^2\theta_W(m_Z)_{\mMS}\nn \\
&+& \frac{\alpha(m_Z)}{4\pi \hat{s}^2} - \frac{3\alpha(m_Z)}{32\pi \hat{s}^2 \hat{c}^2}(1-4 \hat{s}^2)[1+ (1-4\hat{s}^2)^2]\nn \\
&+& F_1(y,Q^2) + F_2(y,Q^2)\Big \},
\eea
where $y=Q^2/s$, $\sqrt{s}$ is the center of mass energy, $\hat{s} \equiv \sin \theta_W(m_Z)_{\mMS}$, and $\hat{c} \equiv \cos \theta_W(m_Z)_{\mMS}$.
The overall factor of $\rho =1 + \cal{O}(\alpha)$  arises from radiative 
corrections~\cite{Marciano:1980pb,Czarnecki:1995fw} to $G_F$, which is defined through the muon decay 
process. 
\begin{figure}[htp]
\noindent
\begin{minipage}{10.cm}
\begin{center}
\begin{tabular}{llll}
\includegraphics[width=3cm, trim = 20mm 140mm 20mm 140mm]{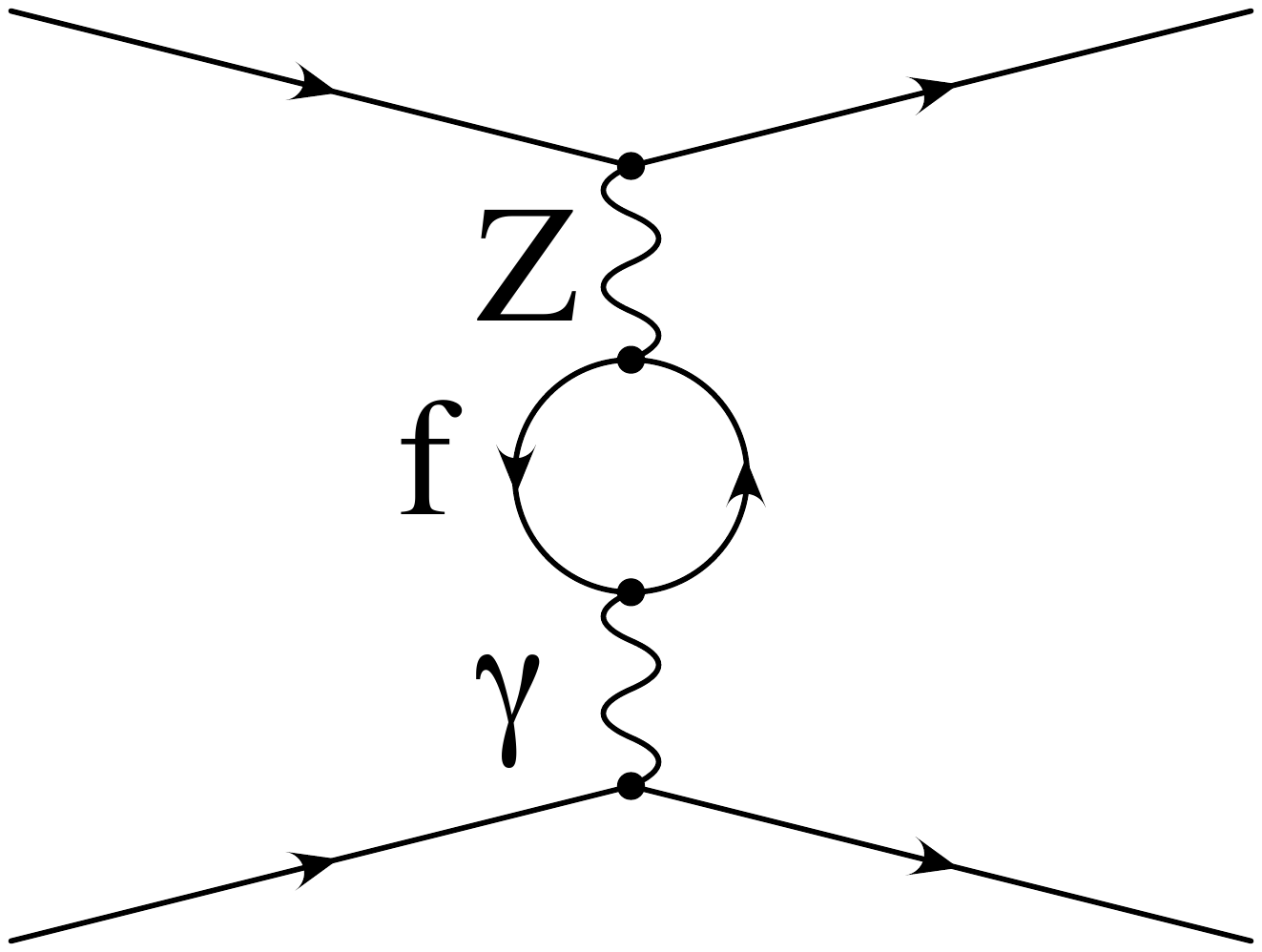}&
\includegraphics[width=3cm, trim = 20mm 140mm 20mm 140mm]{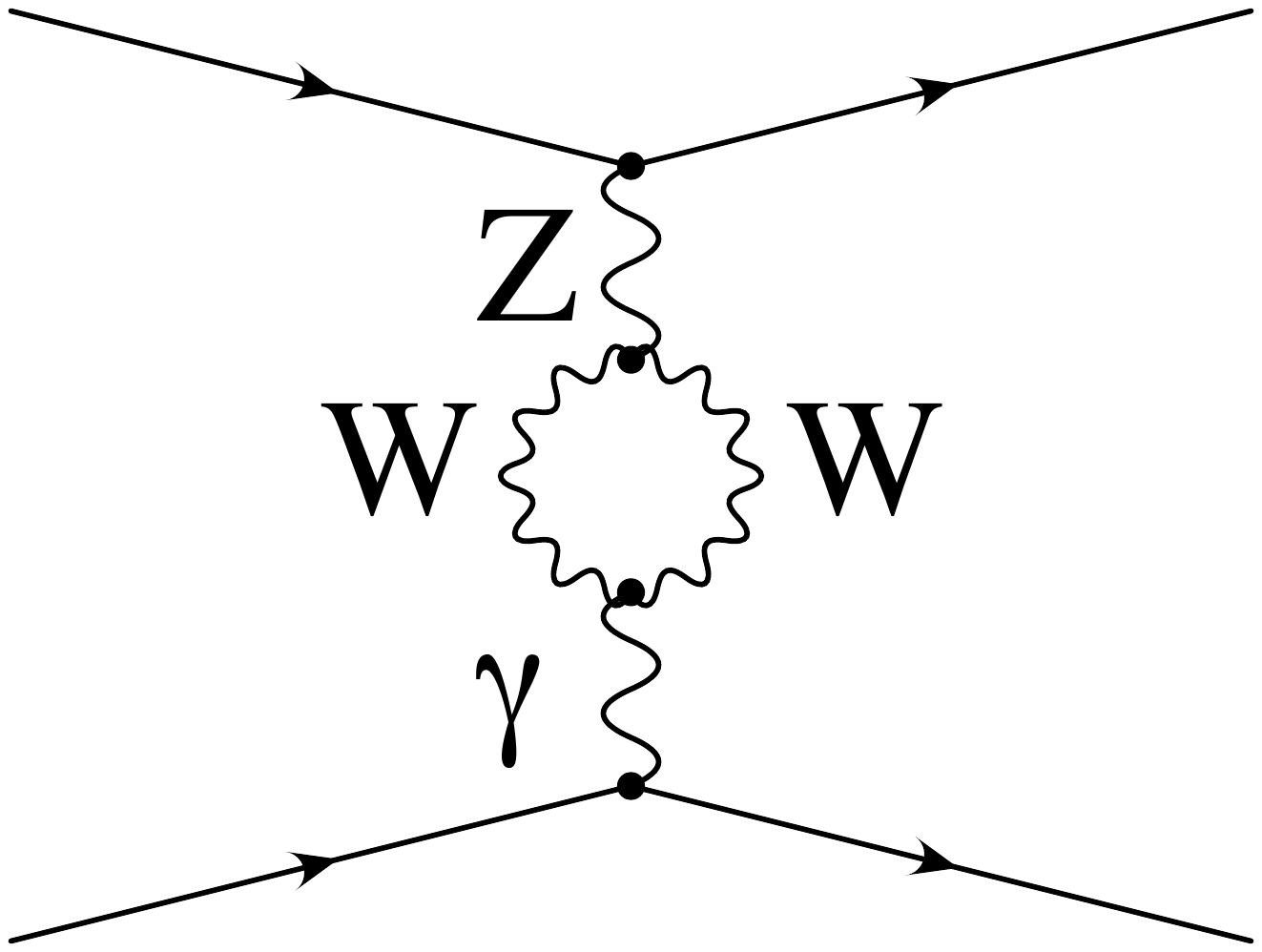}&
\includegraphics[width=3cm, trim = 20mm 140mm 20mm 140mm]{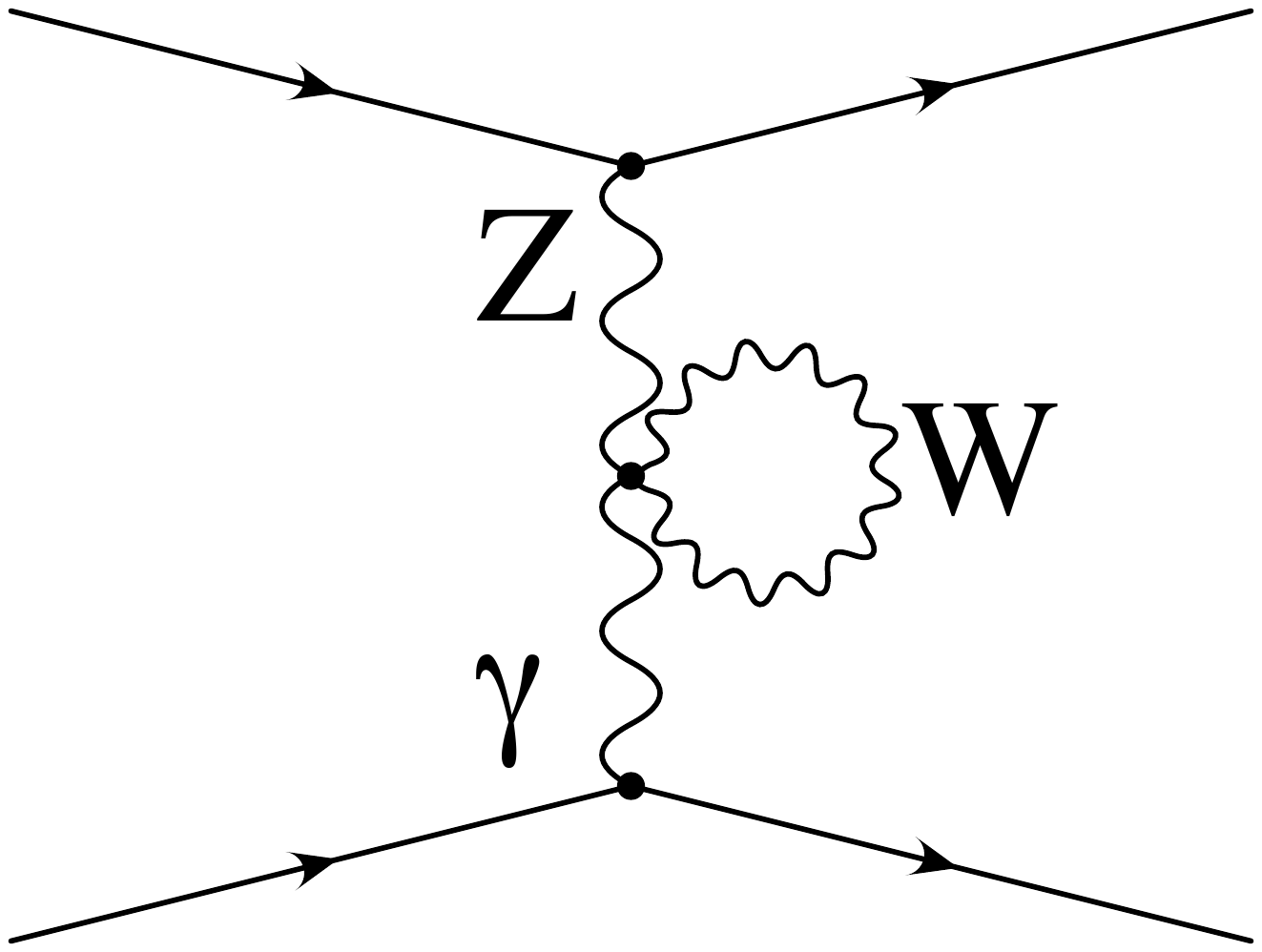}&
\includegraphics[width=3cm, trim = 20mm 140mm 20mm 140mm]{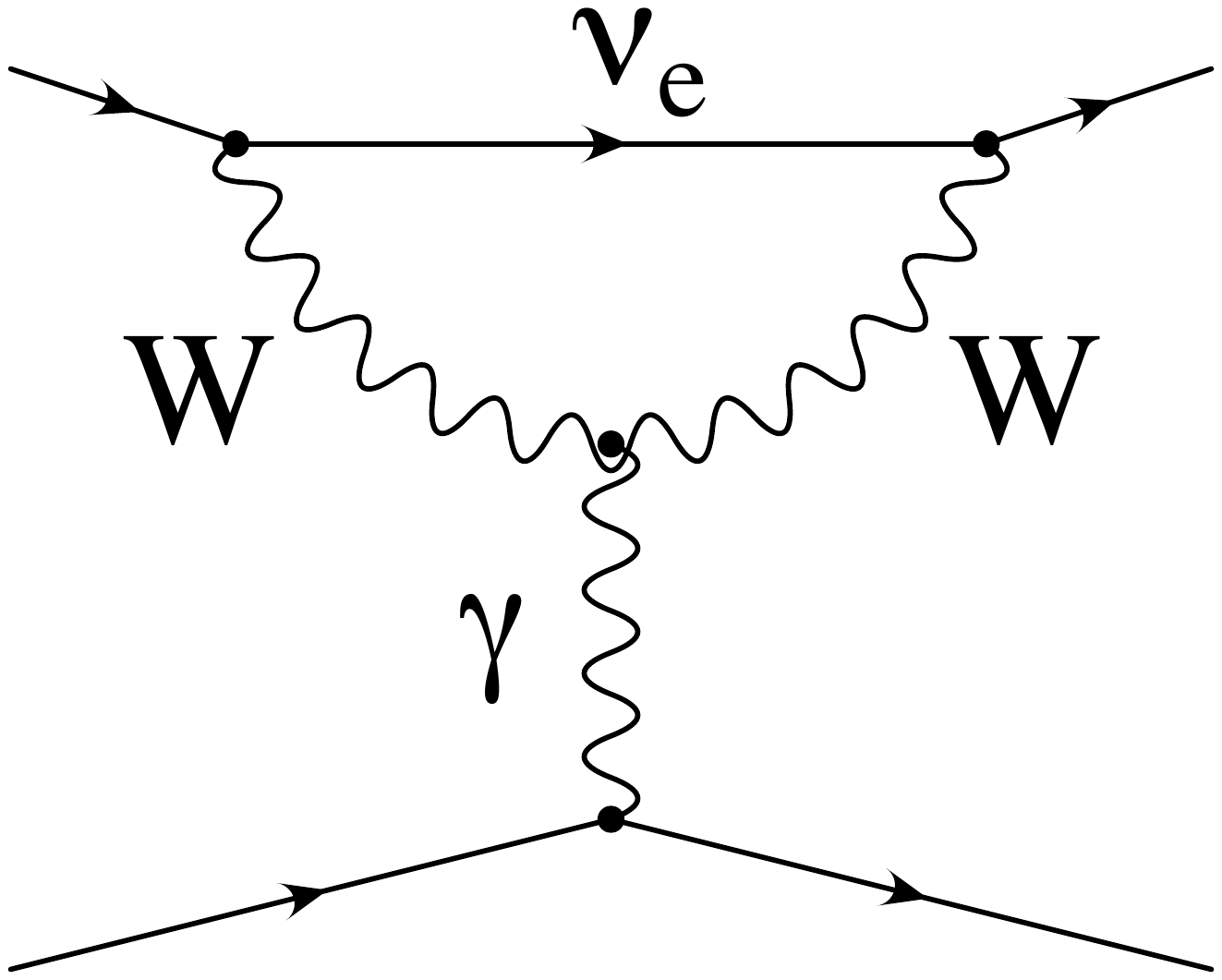}
\end{tabular}
\end{center}
\end{minipage}

\vspace*{15mm}
\noindent 
\caption{ $\gamma-Z$ mixing diagrams and  $W$-loop contribution to
the anapole moment for parity-violating elastic electron scattering (reproduced from Ref.~\cite{czarnecki00})}
\label{feynmolrad}
\end{figure}
The two terms in the second line of 
Eqn~(\ref{moller-radiative}) arise from $WW$ and $ZZ$ box diagram contributions respectively. The $WW$ box
correction gives a $\sim 4\%$ enhancement to the asymmetry. The $ZZ$ box contribution, however, is suppressed
by $(1-4\sin^2\theta_W)$ and gives only a $\sim 0.1\%$ correction. The $F_1(y,Q^2)$ term \cite{Czarnecki:1995fw}
includes  box, external leg, and vertex corrections involving  at least one photon. The dominant effect,
however, arises from the $\gamma-Z^0$ vacuum polarization and anapole moment contributions 
(see Fig.~\ref{feynmolrad}) that are encoded in $\kappa(0)$.  
The effective weak mixing angle at $Q^2=0$ in terms of
the ${\mMS}$ value at the $\zzero$ pole is defined in terms of $\kappa(0)$ as
\bea
\label{sinkappa}
\sin^2\theta_W(0) &=& \kappa(0) \sin^2\theta_W(M_Z)_{\mMS},
\eea
corresponding to Eq.(\ref{thetaQ2}) evaluated at $Q^2=0$. However, the experiment is conducted at finite $Q^2$ and the corresponding finite-$Q^2$  vacuum polarization  effects are contained in $F_2(y,Q^2)$ which is very 
small~\cite{Czarnecki:1995fw}. 

A purely perturbative one-loop calculation gives
\bea
\label{kappa0}
\kappa(0) &=& 1- \frac{\alpha}{2\pi \hat{s}^2} \Big \{ \frac{1}{3} \sum_f (T_{3f} Q_f - 2 \hat{s}^2 Q_f^2) \ln \frac{m_f^2}{m_Z^2}  \nn \\
&-& \Big ( \frac{7}{2}\hat{c}^2 + \frac{1}{12}\Big )\ln \hat{c}^2 + \Big (\frac{7}{9}-\frac{\hat{s}^2}{3} \Big )\Big \},
\eea
where the sum in the first line is over the quark and lepton flavors.
However, it is known that at $Q^2=0$ the light quark contribution to the $\gamma-Z^0$ vacuum polarization is 
non-perturbative and must be estimated using a dispersion relation that relates these effects to data on 
$e^+e^-\to \textit{hadrons}$. The result of such an 
analysis~\cite{Marciano:1982mm, marciano93, Czarnecki:1995fw} leads to the
replacement (for the quark contribution)
\bea
\frac{1}{3}\sum_f (T_{3f} Q_f - 2 \hat{s}^2 Q_f^2) \ln \frac{m_f^2}{m_Z^2} \to -6.88 \pm 0.06\label{vacpol}
\eea
in Eqn.~(\ref{kappa0}), where the error in Eqn.~\ref{vacpol} has been updated by the analysis presented
in Ref.~\cite{Erler:2004in}, which we discuss in detail in the next section. 
It is these hadronic vacuum polarization effects that are primarily responsible for the large $\sim 40\%$ NLO correction to the asymmetry. Coupled with anapole moment effects, they lead to the effective weak mixing angle 
$\sin^2\theta_W(0)$ in Eq.(\ref{sinkappa}) to differ from $\sin^2\theta_W(m_Z)_{\mMS}$ by 3\%, referred to
as the ``running" from $Q^2\sim m_Z^2$ to $Q^2\ll m_Z^2$.

After accounting for the one-loop effects discussed above, one can now define the electron's weak charge
$Q_W^e$ in the limit $E$ and $Q^2\to 0$ as a static property of the electron, with the value quoted
in Eqn.~\ref{eqn:newweak} for $m_H = 126$ GeV.
Efforts are underway \cite{Aleksejevs:2011de,Aleksejevs:2012sq} towards completing the full next-to-next-to-leading order (NNLO) calculation of electroweak radiative corrections to the asymmetry, which are expected
to shift $Q_W^e$ possibly at the 1--2\%\ level. 
Such efforts are essential in the context of the ultra-precise  MOLLER proposal.

\subsection{Running of the Weak Mixing Angle}

An analysis similar to the one presented above must be carried out for any weak neutral current experiment 
that aims to
measure $\sstw$ better than 1\%\ at $Q^2\ll m_Z^2$; careful consideration must be given to the dynamics that lead to the running of $\sstw$. 
In general, perturbative corrections enhanced by large logarithms of $\sim m_Z^2/Q^2$ (or $m_Z^2/m_f^2$, where $m_f$ is some light fermion mass) can significantly affect a reliable interpretation of low $Q^2$ measurements.
For example, the $\sim 40\%$ reduction in the weak
charge of the electron $Q_W^e$, and thus $A_{PV}$ in M\o ller scattering~\cite{Czarnecki:1995fw},
is due to the replacement (as discussed in Secs.~\ref{seclowqsin}, \ref{secE158} and~\ref{mollerrunning})
\bea
\label{repl}
1- 4\sin^2\theta_W(m_Z)_{\mMS} \to 1- 4\kappa (0)\sin^2\theta_W(m_Z)_{\mMS}.
\eea
\noi $\kappa(0)$ encodes the radiative corrections from $\gamma-Z^0$ mixing and some 
anapole moment effects, motivating the definition of an effective weak mixing angle
\bea
\label{kap0}
\sin^2\theta_W(0) \equiv \kappa (0)\sin^2\theta_W(m_Z)_{\mMS},
\eea
which simply corresponds to Eq.(\ref{thetaQ2}) evaluated at $Q^2=0$.   The quantity $\sin^2\theta_W(0)$ 
incorporates a set of universal radiative corrections that also affect other low energy measurements such as 
APV and Qweak. The perturbative one loop result  \cite{Czarnecki:1995fw} for $\kappa(0)$ was  given in 
Eqn.~\ref{kappa0} and non-perturbative effects are incorporated by the replacement in Eqn.~\ref{vacpol}. 
Note that the one loop result for $\kappa(0)$ contains large logarithms of $m_Z^2/m_f^2$ (at finite $Q^2$ there will 
also be large logarithms of $m_Z^2/Q^2$) that can spoil convergence. Higher precision extractions of $\sstw$ 
at low $Q^2$ require a resummation of such large logarithms, bringing the theory under better control and 
facilitating a more precise interpretation of measurements.

A well-known way to incorporate resummation of large logarithms is to work in the $\mMS$-scheme for $\sstw$
as defined in Eqn.~\ref{eqseventeen}. It uses a well-defined gauge independent subtraction scheme to
remove divergent terms arising in  calculations of quantum corrections that use dimensional regularization. This
subtraction scheme induces a logarithmic dependence on the renormalization scale $\mu$ which is governed by a
renormalization group (RG) equation. Choosing $\mu^2 \sim Q^2$ of the process avoids the appearance of large
logarithms in $m_Z^2/Q^2$. The $\mMS$ scheme also employs threshold matching to avoid  large logarithms in
$m_Z^2/m_f^2$ when $\mu\gg m_f$ or $\mu\ll m_f$. Crossing the particle mass threshold from above, the
corresponding particle is integrated out and the running below continues within an effective theory without this
particle. These threshold matchings manifest themselves as discontinuities in the weak mixing angle
running.

In the $\mMS$ scheme, the quantity of interest for low energy experiments is $\sin^2\theta_W (0)_{\mMS}$ 
corresponding $\mu=0$ in Eqn.~\ref{eqseventeen}. It is defined in terms of the $\zzero$-pole value of the weak
mixing angle as $\sin^2\theta_W (0)_{\mMS}=\kappa(0)_{\mMS}\sin^2\theta_W (m_Z)_{\mMS}$, where the quantity $\kappa(0)_{\mMS}$ is obtained by solving the renormalization group equation  between $\mu= m_Z$ and $\mu=0$. More generally, 
\bea
\label{eq:szero}
\sin^2\theta_W (Q^2)_{\mMS} = \kappa(Q^2,\mu)_{\mMS} \> \sin^2\theta_W (\mu^2)_{\mMS},
\eea
 gives the relation between the weak mixing angle at some fixed $Q^2$ in terms of its value at an arbitrary scale 
 $\mu$. Note that the product $\kappa(Q^2,\mu)_{\mMS} \> \sin^2\theta_W (\mu^2)_{\mMS}$ in 
 Eqn.~\ref{eq:szero} is independent of the scale $\mu$. This allows one to choose $\mu^2=Q^2$  (along with 
threshold matching), effectively moving large logs from $ \kappa(Q^2,\mu)_{\mMS}$ into 
$\sin^2\theta_W(\mu^2)_{\mMS}$ so that resummation can be performed using the RG evolution equation of 
 $\sstw$. On the other hand, choosing $\mu^2 =m_Z^2$ introduces large logarithms  of $Q^2/m_Z^2$ in 
 $\kappa_{\mMS}(Q^2,\mu=m_Z)$ spoiling the convergence of perturbation theory. In Ref.~\cite{Erler:2004in}, a solution to the RG equation of $\sin^2\theta_W (\mu)_{\mMS}$, for evolution between  scales $\mu_0$ and $\mu$ without crossing any particle mass thresholds, was given to be
\bea
\label{resum}
\sin^2\theta_W(\mu)_{\mMS} &=& \frac{\alpha(\mu)_{\mMS}}{\alpha(\mu_0)_{\mMS}} \sin^2\theta_W(\mu_0)_{\mMS}+\lambda_1 \Big [ 1- \frac{\alpha(\mu)}{\alpha(\mu_0)}\Big ] \nn \\
&+& \frac{\alpha(\mu)}{\pi} \Big [ \frac{\lambda_2}{3}\ln \frac{\mu^2}{\mu_0^2} + \frac{3\lambda_3}{4}\ln \frac{\alpha(\mu)_{\mMS}}{\alpha(\mu_0)_{\mMS}}+\tilde{\sigma}(\mu_0) - \tilde{\sigma}(\mu)\Big ].
\eea
In the above equation, $\lambda_{1,2,3}$ are numerical coefficients that take on different values depending on the 
range $(\mu_0,\mu)$. This solution resums leading logs ${\cal O}(\alpha^{n}\ln ^n\frac{\mu}{\mu_0})$, 
next-to-leading logs  ${\cal O}(\alpha^{n+1}\ln ^n\frac{\mu}{\mu_0})$ and 
${\cal O}(\alpha \alpha_s^{n}\ln ^n\frac{\mu}{\mu_0})$, next-to-next-to-leading logs 
${\cal O}(\alpha \alpha_s^{n+1}\ln ^n\frac{\mu}{\mu_0})$, and next-to-next-to-next-leading logs 
${\cal O}(\alpha \alpha_s^{n+1}\ln ^n\frac{\mu}{\mu_0})$. Non-perturbative effects arise from the contribution of light 
quark loops in self-energy $\gamma-Z^0$ mixing diagrams when $\mu \sim \Lambda_{QCD}$. These non-perturbative
effects are incorporated in Eqn.~\ref{resum} through the non-perturbative effects in the evolution of 
$\alpha(\mu)_{\mMS}$ and in the $\tilde{\sigma}(\mu_0), \tilde{\sigma}(\mu) $ terms. These non-perturbative effects
contribute an uncertainty in the extraction of $\sin^2\theta_W(0)_{\mMS}$ below the $10^{-4}$ level.

The value of $\sin^2\theta(0)_{\mMS}$, in terms of $\sin^2(m_Z)_{\mMS}$, can be obtained by using 
Eqn.~\ref{resum} combined with threshold matchings to evolve between the scales $\mu=m_Z$ and $\mu=0$. 
It was shown in Ref.~\cite{Erler:2004in} that  the solution to the $\mMS$ RG evolution, expanded to one-loop order is
\bea
\kappa(0)_{\mMS} =\kappa(0) + \frac{2\alpha(m_Z)}{9\pi\hat{s}^2}= \kappa(0)^{PT}= 1.03232\pm 0.00029,
\eea
\noi where non-perturbative effects have been included. The uncertainty has been improved by an order
of magnitude compared to previous estimates. 
The quantity $\kappa(0)^{PT}$ \cite{Ferroglia:2003wa}, is related to $\kappa(0)$ by the inclusion of additional 
so-called ``pinch-parts" of one-loop vertex and box graphs to make it process-independent and intrinsically gauge-invariant. In the $\mMS$ scheme, these additional pinch terms arise \cite{Erler:2004in} from threshold matching corrections at $\mu=m_W$.  Thus, working in the $\mMS$ scheme allows one to reproduce the known one-loop result while allowing for the inclusion of the leading higher order corrections through resummation. 

In Fig.~\ref{fig:run}, we show the running of $\sin^2\theta_W(Q^2)$ and $\sin^2\theta_W(Q^2)_{\mMS}$ for comparison. Based on the work described above and the prediction for $\sstw$ discussed 
Sec.~\ref{intro} using fundamental SM input parameters including $m_H$, we obtain:
\bea
\sstw(m_Z)_{\mMS}= 0.23124(6)\rightarrow \sstw(0)_{\mMS} = 0.23871(9).
\eea

\subsection{Weak Charge of the Proton}
\label{secqweaktheory}

\begin{figure}[tb]
\includegraphics[height=2in,width=4in]{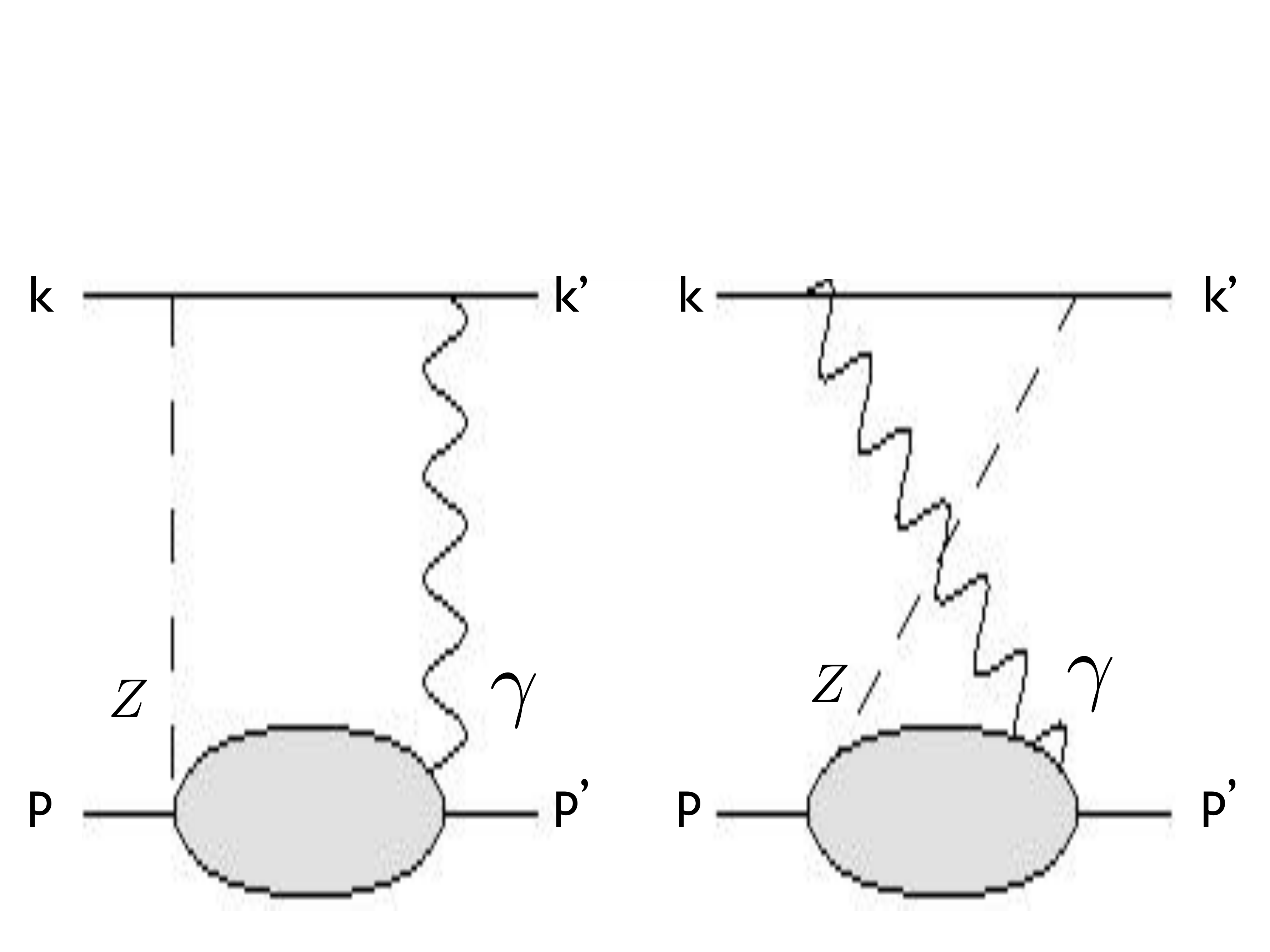}
\caption{Box graphs contributing to $\Box_{\gamma Z}$ in elastic electron-proton scattering (reproduced
from Ref.~\cite{Gorchtein:2008px})}
\label{fig:box}
\end{figure}

One can extract $Q_W^p$ from the measurement of $A_{PV}$ in electron-proton scattering. 
However, that requires considerations beyond the perturbative approach used for M\o ller scattering
(Sec.~\ref{mollerrunning}). 
In particular, one must address hadronic physics that induces an energy dependence~\cite{Gorchtein:2008px} 
to the $\gamma-\zzero$ box diagram (Fig.~\ref{fig:box}). 
A definition of the weak charge that isolates such effects is~\cite{Gorchtein:2011mz}
\bea
\label{Qwpdef}
A_{PV} &=&  -\frac{G_F Q^2 }{4\sqrt{2}\pi \alpha} \frac{W^{PV}}{W^{EM}},\qquad Q_W^p = \lim_{Q^2\to 0} \frac{W^{PV}}{W^{EM}} \Big |_{E=0},
\eea
where we have written the asymmetry in terms of the general   response functions $W^{EM}, W^{PV}$ which depend on the electromagnetic and weak nucleon form factors. The conditions $E=0, Q^2=0$ in  the definition of $Q_W^p$, ensure that $Q_W^p$ can be interpreted as a static property of the proton, independent of  kinematics.

The tree-level SM value of $Q_W^p= 2 (2 C_{1u} + C_{1d})= 1- 4\sin^2\theta_W$, corresponding to the sum of the 
weak charges of the valence quarks in the proton, receives corrections from perturbative radiative effects and
non-perturbative hadronic effects. The asymmetry $A_{PV}$ can be written as
\bea
\label{APVbox}
A_{PV} &=& -\frac{G_F Q^2}{4\sqrt{2} \pi \alpha} \Big [ \rho_{ep}(1-4 \sin^2\theta_W(0)_{\overline{MS}}) \nn \\
\>\>\>\>&+&  \mbox{Re }\Box_{WW} +  \mbox{Re }\Box_{ZZ} +  \mbox{Re }\Box_{\gamma Z}  \Big ] -\frac{G_F Q^2}{4\sqrt{2} \pi \alpha}  B(Q^2)  ,
\eea
where $ \rho_{ep}=1+\cal{O}(\alpha)$ radiative corrections (absorbing universal and process-dependent terms explicitly defined
in Eqn. 12 of Ref.~\cite{Gorchtein:2011mz}) due to the normalization of the weak neutral current $ep$ amplitude relative to the charged current muon decay amplitude used to define $G_F$, and 
$\Box_{WW},\Box_{ZZ}$, and $\Box_{\gamma Z}$ are contributions from the two-boson box graphs in Fig. (\ref{fig:box}). The remaining contribution $B(Q^2)$ parametrizes proton structure
at low $Q^2$ and vanishes in the forward limit ($B(Q^2)\to 0$  as $Q^2\to 0$). Comparing 
Eqns.~\ref{Qwpdef}--\ref{APVbox} gives
\bea
\label{Qwpdef-1}
Q_W^p &=&  \Big [ \rho_{ep}(1-4 \sin^2\theta_W(0)_{\mMS}) \nn \\
\>\>\>\>&+&  \mbox{Re }\Box_{WW} +  \mbox{Re }\Box_{ZZ} +  \mbox{Re }\Box_{\gamma Z}  \Big ] \Bigg |_{E=0,\>Q^2\to 0}.
\eea
%Eqs. (\ref{APVbox}) and (\ref{Qwpdef-1}) show the correct relationship between the proton weak charge and the measured asymmetry.  

All the  box graphs appearing in Eqn~(\ref{APVbox}) are ultraviolet finite.
The $\Box_{WW}$ and $\Box_{ZZ}$ box graphs ~\cite{Marciano:1983wwa, Marciano:1982mm, Erler:2003yk} give 
perturbative corrections of  $\sim 26\%$ and $\sim 3\%$ to $Q_W^p$ respectively, independent of the electron
energy since loop momenta of order $M_Z$ dominate.   Remarkably, these corrections nearly cancel the reduction
in $Q_W^p$ due to the effect of 
$\kappa(0)$ (Eqn.~\ref{eq:szero}), which makes it seem like $Q_W^p$ does not run with $Q^2$, in stark 
contrast to $Q_W^e$. 

The calculation of the $\Box_{\gamma Z}$ contribution is more complicated since it is sensitive to small momentum scales and non-perturbative long distance physics. 
%%%OMIT%%%%
\OMIT{\textbf{The $\gamma \gamma$ box graphs \cite{Carlson:2007sp} ... (discuss this more)... Also, discuss the earlier calculations of  $\>\Box_{\gamma Z}$ that did not use dispersion relations (Marciano, Sirlin). }} 
%%%END-OMIT%%%%%
Pinning down the size of the correction and theoretical uncertainty of the $\Box_{\gamma Z}$ contribution 
has been the subject of active research for almost  three decades and continues even today.
This contribution can be written as the sum of two terms
\bea
\label{boxphz}
\Box_{\gamma Z} &=& \Box_{\gamma Z_A} + \Box_{\gamma Z_V},
\eea
corresponding to the $\zzero$-electron axial-vector ($g_A^e$) and vector ($g_V^e$) coupling contributions
respectively. The first calculation of  $\Box_{\gamma Z}$ was carried 
out~\cite{Marciano:1982mm,Marciano:1983wwa} in the context of atomic parity violation where the electron energy
$E\approx 0$. A cancellation between the box and crossed box graphs leads to a negligible contribution from 
$\Box_{\gamma Z_A}$. The second and dominant contribution is
\bea
\label{msbox}
\mbox{Re } \Box_{\gamma Z_V} = \frac{5\alpha(m_Z)_{\mMS}}{2\pi}\>g_V^e\>\Big [ \ln \frac{m_Z^2}{\Lambda^2} + C_{\gamma Z}(\Lambda)\Big ]
\eea
suppressed by the weak vector coupling of the electron $g_V^e=1-4\sin^2\theta_W(m_Z)_{\mMS}$. In 
Eqn.~\ref{msbox}, $\Lambda \sim 1$ GeV, is a hadronic cutoff that separates the perturbative and 
non-perturbative contributions. The first term arises from a perturbative contribution from  loop momenta greater 
than $\Lambda$ and the second term is the remaining non-perturbative contribution estimated to be 
 $C_{\gamma Z}(\Lambda)=3/2\pm 1$ for $\Lambda=m_\rho$. This estimate was based on the Born approximation
where the elastic proton intermediate  state dominates the hadronic effects. The perturbative contribution in 
Eqn.~\ref{msbox} was recomputed~\cite{Erler:2004in,Erler:2003yk,Musolf:1990ts}, confirming the original result. 

Recently, the $\Box_{\gamma Z}$ contribution was reexamined in the context of the kinematics of the Qweak experiment. It was shown \cite{Gorchtein:2008px} that there is a contribution, not considered in previous analyses, that grows with the incident electron energy. In the forward limit $Q^2\to 0$, a dispersion relation  relates the real and imaginary parts of the $\Box_{\gamma Z}$ contribution as
\bea
\label{dispersion}
 \mbox{Re }\Box_{\gamma Z_A}(E) &=& \frac{2E}{\pi} \int_{\nu_\pi}^\infty \frac{d\nu '}{\nu'^2- E^2} \mbox{Im }\Box_{\gamma Z_A}(\nu'), \nn \\
 \mbox{Re }\Box_{\gamma Z_V}(E) &=& \frac{2}{\pi} \int_{\nu_\pi}^\infty \frac{\nu ' d\nu '}{\nu'^2- E^2} \mbox{Im }\Box_{\gamma Z_V}(\nu'),
\eea
and the imaginary parts are given in terms of the PVDIS $\gamma-Z$ interference structure functions $F_{1,2,3}^{\gamma Z}(x,Q^2)$ as \cite{Gorchtein:2008px,Sibirtsev:2010zg,Rislow:2010vi,Gorchtein:2011mz}
\bea
\label{dispersion-2}
\frac{ \mbox{Im }\Box_{\gamma Z_A}}{\alpha \>g_A^e}&=& \int_{W_\pi^2}^s \frac{dW^2}{(s-M^2)^2}\int_0^{Q^2_{\mathrm{max}}}\frac{dQ^2}{1+\frac{Q^2}{M_Z^2}}\Big [F_1^{\gamma Z} + \frac{s(Q^2_{\mathrm{max}}-Q^2)}{Q^2(W^2-M^2+Q^2)}F_2^{\gamma Z} \Big ] \nn \\
\frac{ \mbox{Im }\Box_{\gamma Z_V}}{-\alpha \>g_V^e }&=& \int_{W_\pi^2}^s \frac{dW^2}{(s-M^2)^2}\int_0^{Q^2_{\mathrm{max}}}\frac{dQ^2}{1+\frac{Q^2}{M_Z^2}}\Big [\frac{2(s-M^2)}{W^2-M^2+Q^2}-1\Big ]F_3^{\gamma Z} %\nn \\
 \eea
 where $x=Q^2/(2p\cdot q)$, $W^2=(p+q)^2$, $W_\pi^2=(M+\pi)^2$, $\nu_\pi=(W^2_\pi-M^2)/(2M)$, $Q^2_{\mbox{max}}=(s-M^2)(s-W^2)/s$, and $p$ is the initial proton momentum.
 The explicit overall factor of the electron energy $E$ in $ \mbox{Re }\Box_{\gamma Z_A}$, seen in 
 Eqn.~\ref{dispersion},  is the origin of the new electron energy dependence in  $\Box_{\gamma Z}$. 
 Note that while the $\Box_{\gamma Z_V}$ contribution is suppressed by $g_V^e$, no such suppression exists for 
 $\Box_{\gamma Z_A}$, resulting in a significant energy-dependent correction to the Qweak asymmetry.
 
As seen in Eqns.~\ref{dispersion}--\ref{dispersion-2}, estimating the size of the $\Box_{\gamma Z}$ contribution 
requires knowledge of the $F_{1,2,3}^{\gamma Z}$ structure functions over a wide range of kinematics. This range
can be classified into three regions: (i) elastic ($W^2=M^2$), (ii) resonance ($\>W_\pi^2 \leq W^2 \lesssim 4 $GeV$^2$ ), and (iii) deep inelastic
($W^2 > 4$ GeV$^2$). 
Note that since $Q^2=0.026$ GeV$^2$ in Qweak, the structure functions 
cannot be expressed in terms of the leading twist pdfs as is usually done when $Q^2\gg \Lambda_{QCD}^2$.  A
combination of data and modeling of the structure functions is necessary in the three regions for an accurate
estimate  of $\Box_{\gamma Z}$. Moreover, the dispersion relations in 
Eqn.~\ref{dispersion} were derived in the forward limit, leading to an additional $Q^2$ dependence in the box graphs mentioned above which was found~\cite{Gorchtein:2011mz} 
to be small. 

There have been several recent estimates~\cite{Gorchtein:2008px,Sibirtsev:2010zg,Rislow:2010vi,Gorchtein:2011mz,Blunden:2011rd} 
of the size of the $\Box_{\gamma Z}$ contribution and its associated uncertainty in determining $Q_W^p$. 
While this is still an active area of research with some differences between approaches and results, 
the general consensus is that $\Box_{\gamma Z}$ contributes a
$\sim 5-6\%$ correction for $E_{beam}\sim 1$ GeV that must be subtracted to determine $Q_W^p$
as defined in Eqn.~\ref{Qwpdef-1}, with a $\sim2-3\%$ uncertainty; 
the anticipated experimental uncertainty is 4\%. The larger of the uncertainty estimates has a contribution
from model dependence associated with flavor rotations. Auxiliary measurements from current and future
JLab experiments discussed in this review 
will cover a wider range of $F_{1,2,3}^{\gamma Z}$ and further reduce the theoretical uncertainty in 
$\Box_{\gamma Z_A}$. The theoretical uncertainty is already negligible for the P2 proposal (Sec.~\ref{sec:P2}),
with $E_{beam}\sim 0.2$ GeV.

\section{OTHER POTENTIAL FUTURE MEASUREMENTS}
\label{secother}
\subsection{Atomic Parity Violation}

A number of new APV projects are underway.  Two separate initiatives are being pursued on 
heavier atoms to take advantage of the greatly enhanced parity-violating amplitudes at higher values of $Z$. 
At TRIUMF in 
Canada, an experiment is under design to use Fr~\cite{Aubin:2012zz}  atoms while at KVI in the Netherlands, 
Ra$^+$ ions~\cite{Giri:2011zz} are being investigated.  In both designs, 
chains of isotopes could potentially be measured, such that the atomic theory uncertainties cancel 
when ratios of different isotopes are taken.  However, it must be pointed out that the measurement of 
$Q_W$ for a single isotope, properly normalized to the atomic theory, will remain important since it 
has rather different sensitivities to new physics effects. Indeed, there are often scenarios where new physics effects
also tend to cancel when isotope ratios are taken. There are also plans
to measure isotope chains in Dy and Yb~\cite{Tsigutkin:2009zz}, though the atomic theory is more challenging.
These measurements will also be sensitive to the thickness of the radius
of the neutron distribution, which is a subject of considerable	interest in
itself~\cite{Sil:2005tg}.

\subsection{Neutrino Scattering}

Currently, the NuTeV deep-inelastic neutrino scattering result (Eqn.~\ref{eqthirtyfive}) represents the best
neutrino determination of the weak mixing angle ($\sim\pm 0.7\%$). It differs by nearly 3$\sigma$ from SM 
expectations; a situation that requires resolution. The advent of future high intensity neutrino sources, designed
primarily for neutrino oscillation studies, could in principle resolve the NuTeV anomaly. For example, a fine-grained
near-detector at the Fermilab LBNE facility has been suggested as a means of achieving $\pm 0.2\%$ sensitivity
at $Q^2$ values similar to NuTeV~\cite{Mishra:2008nx}.

Low energy neutrino sources such as nuclear power reactors, spallation neutron sources, $\beta-$beams etc
used in conjunction with very massive detectors would be capable of measuring $\sstw(m_Z)_{\mMS}$ at very
low $\langle Q\rangle\sim 1-30$ MeV using $\nu_e$ and $\nu_\mu$ scattering on electrons. Fractional
sensitivities of $\sim\pm 1\%$ on $\sstw$ appear feasible. Unfortunately, to reach the $\pm 0.1\%$ goal
appears very challenging both statistically and systematically. Nevertheless, it is a well-motivated goal,
since at that level they probe many interesting varieties of ``new physics"~\cite{Marciano:2003eq, 
Conrad:2004gw, Rosner:2004yt, deGouvea:2006cb, Agarwalla:2010ty}.

\subsection{Parity Violating Electron Scattering off $^{12}$C}

More than two decades ago, a measurement~\cite{Souder:1990ia} in $\vec{\rm e}-^{12}$C elastic
scattering of $A_{PV}(eC)\propto G_FQ^2\sstw$ at the MIT-Bates laboratory achieved $\pm 25$\%\ precision.
Along the way, techniques were developed that set the stage for parity-violating electron scattering experiments
of the type discussed in this review. 

Today, higher electron currents with better longitudinal polarization combined with a much larger acceptance 
spectrometer~\cite{Souder:1990ie} could potentially improve the statistical figure of merit by $10^4$, making
an asymmetry measurement of $Q_W(^{12}{\rm C})$ to about $\pm 0.3\%$ statistically feasible. While controlling
the sources of spurious false asymmetries at the level required have been developed and are similar to those
of other proposed $A_{PV}$ measurements, controlling normalization errors and in particular the measurement 
of the electron beam polarization will require a detailed study. Assuming a total $\pm 0.3\%$ determination of 
$Q_W(^{12}{\rm C})$ is possible, what can we learn, compared to other $A_{PV}$ measurements, from such 
an effort?

Referring to the discussion in Sec.~\ref{secnewweak} and Eqn.~\ref{eqn:newweak} in particular,
we already pointed out that in terms of  $m_{Z_\chi}$ sensitivity,  a $\pm 0.3\%$ measurement of 
$Q_W(^{12}{\rm C})$ is roughly equivalent to the future MOLLER proposal to determine $Q_W(e)$ to $\pm 2.3\%$. 
It also represents about a factor of 3
improvement over $Q_W({\rm Cs})$ where a 1.5$\sigma$ difference between theory and experiment currently exists.

Perhaps another compelling motivation for a new ultra-precise $Q_W(^{12}{\rm C})$ measurement comes from the 
sensitivity to a light $Z_d$ dark gauge boson~\cite{Davoudiasl:2012ag,Davoudiasl:2012qa} 
introduced in Sec.~\ref{secdarkz}. 
%where the combination of $\gamma-Z_d$ kinetic mixing
%and $\zzero-Z_d$ mass mixing gives rise to a new source of ``dark parity violation" in low $Q^2$ experiments. 
Indeed, explaining the $(g-2)_\mu$ $3.6\sigma$ discrepancy with 
20 MeV $\lsim m_{Z_d}\lsim$ 50 MeV and $\epsilon\approx 2\times 10^{-3}$ 
$\gamma-Z_d$ mixing, as well as accommodating the $1.5\sigma$ discrepancy in $Q_W$(Cs)
requires an $X(Q^2)$ corresponding to 
\bea
\Delta\sstw(Q^2) \approx -0.003(2)\frac{m_{Z_d}^2}{Q^2+m_{Z_d}^2}
\eea
which is to be compared with a $\pm 0.0007$ experimental sensitivity of a $\pm 0.3\%$ $A_{PV}(^{12}{\rm C})$
measurement. 

At the Mainz-MESA facility, the measurement might be feasible using the same 
concept as that for the P2 proposal (Sec.~\ref{sec:P2}). With the projected technical capabilities for
other proposed $A_{PV}$ initiatives, a series of three simultaneous $\pm 0.3\%$ measurements  
with $100\lesssim Q\lesssim150$ MeV at $E_{\rm beam} \sim 200$ MeV and a second series of 
$\pm 0.4\%$ measurements with $60\lesssim Q\lesssim100$ MeV at $E_{\rm beam} \sim 140$ MeV could be contemplated. While the statistics
would be achievable in two 1-year runs, the systematic control of normalization errors merits
a detailed study. 

Getting to a lower $Q$ than 60 MeV would be challenging without 
new experimental technologies. Nevertheless, a program of measurements could become quite compelling 
depending on the results of ongoing $A_{PV}$ and dark matter initiatives, or if the 
50--100 MeV mass range for $Z_d$ becomes important to explore
Together, a factor of 3 overall
improvement compared to the existing Cs APV measurement, and the low $Q^2$ sensitivity to $X(Q^2)$ and 
``dark $Z_d$" effects may be enough to motivate a new, much higher sensitivity $A_{PV}(^{12}{\rm C})$ program.

\subsection{Weak Mixing Angle at an Electron-Ion Collider}

The design of a new experimental facility known as the Electron Ion Collider (EIC) is under study as the next logical
step in the study of QCD in nuclear matter~\cite{Deshpande:2012bu}. 
One operational mode will involve high energy collisions of highly
polarized electrons with polarized $^1$H, $^2$H and $^3$He with $20\lesssim\sqrt{s}\lesssim 150$ GeV and
luminosity $\sim 10^{33-34}\mathrm{cm}^2\mathrm{s}^{-1}$.  The collider environment and the envisioned
hermetic detector package at high luminosity will allow precision $A_{PV}$ measurements over a wide 
kinematic range with little uncertainty from limited knowledge of pdfs and negligible impact of higher-twist effects.

\begin{figure}
\begin{center}
  \includegraphics[width=0.95\columnwidth]{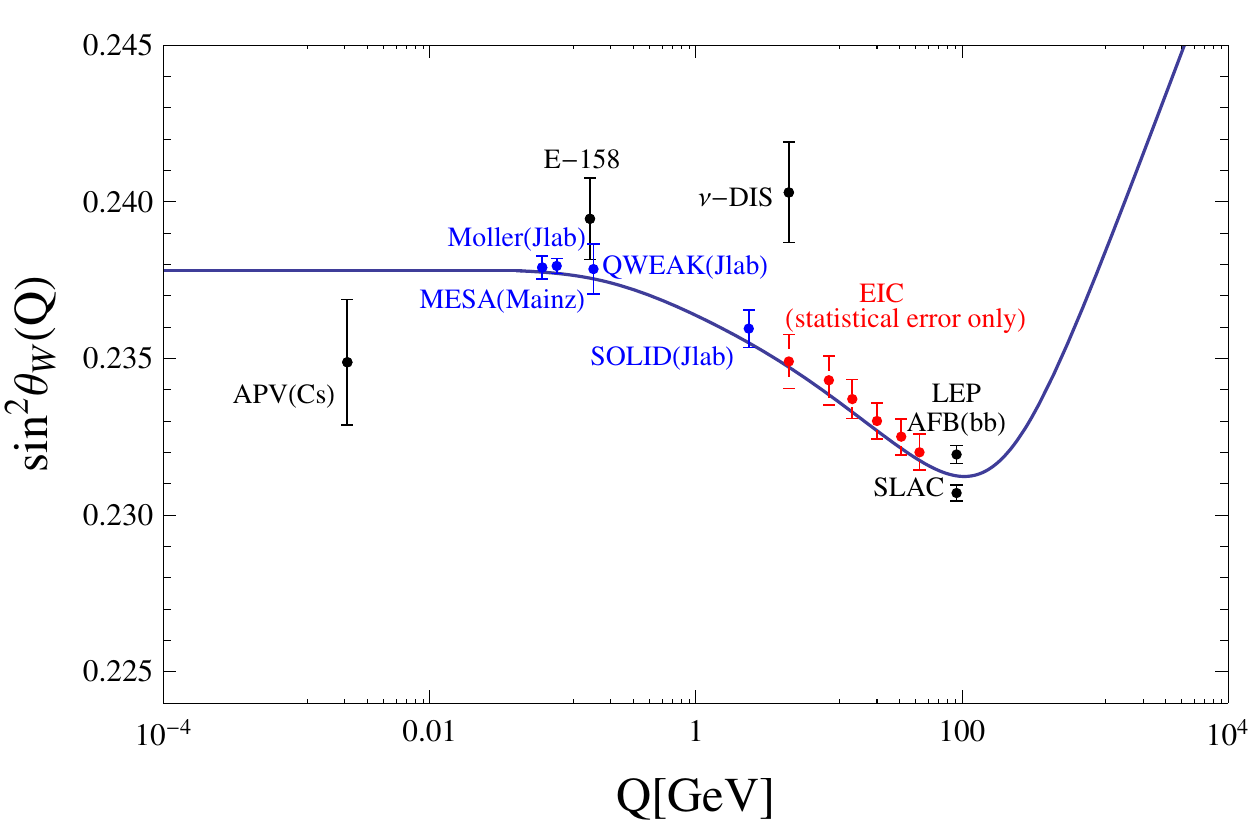}
\end{center}
%\vspace{3in}
\caption{Current and future $\sin^2\theta_W$ measurements. 
 The black points points are published results (Sec.~\ref{secpast}), the blue points are projections for projects
discussed in Sec.~\ref{secprogram}, and the red points are projected
for the EIC, $\sqrt{s} = 140$ GeV, 200 fb$^{-1}$.}
\label{sinthetarunning}
\end{figure}

By mapping $A_{PV}$
as a function of $Q^2$ and the fractional energy loss of the scattered electron $y$ (something that is 
very challenging to do in fixed target experiments), a clean separation of two linear combination of
couplings namely $2C_{1u}-C_{1d}$ and $2C_{2u}-C_{2d}$ will become feasible as a function of $Q^2$.
Thus, at the highest luminosities and center-of-mass energies envisioned 
at the EIC, very precise measurements of these combinations can be achieved at a series of $Q^2$ values,
allowing a series of precision $\sstw$ extractions. 
Figure~\ref{sinthetarunning} shows the projected uncertainties on the weak mixing angle extracted
from such a dataset~\cite{Boer:2011fh}, for a center of mass energy of 140 GeV and an 
integrated luminosity of 200 fb$^{-1}$.

\section{SUMMARY AND OUTLOOK}
\label{secconclude}

Precision measurements of electroweak parameters have played an important role in confirming the SM
and probing for ``new physics" effects. The recent (tentative) discovery of the Higgs boson at the LHC, with 
$m_H\approx 126$ GeV completes the elementary particle spectrum of the minimal SM and allows even more
refined predictions and comparisons between theory and experiment. 

Already, determinations of $m_W$ and $\sstw(m_Z)_{\mMS}$ (at the $\zzero$ pole) test SM expectations 
at better than $\pm 0.1$\%\ and find no glaring evidence for discrepancies (although $m_W$ is somewhat high
and $\sstw(m_Z)_{\mMS}$ measurements have a broader spread than one would like). Those tests confirm the 
SM at its quantum loop level and constrain ``new physics" appendages to it; examples include fourth generation
fermions, technicolor, supersymmetry, GUTs etc.

Low $Q^2$ SM observables, such as weak neutral current charges $Q_W$, can also be precisely computed
and compared with measurements. Results from classic measurements such as Cs APV and SLAC E158
are currently at about the $\pm 0.5-1\%$ sensitivity for $\sstw(Q^2\approx 0)$. At that level, they test quantum
loops and confirm the anticipated running of $\sstw(Q^2)$ by about 3\%\ as it evolves from $Q^2\approx 0$
to $m_Z^2$. In addition, those measurements play a special role in constraining $Z'$, leptoquark models, 
and generic 4-fermion contact interactions. 

There is rich physics in the radiative corrections that leads for example to the remarkable numerical difference between the electron and the proton renormalized weak charges. This has implications for the design,
feasibility and systematic error propagation for precision experiments. The difference also serves to emphasize the value of studying
the SM with high precision in as many reactions as possible.   A priori,
one cannot know where new physics may be observed; physics that
can be described at low energies by new contact interactions or physics that appears in vacuum polarization
or box diagrams at the quantum loop level. 

Now a new generation of polarized electron scattering experiments are on the horizon. Qweak
at JLab has completed data collection and is in the analysis stage. It will improve the low $Q^2$ determination
of $\sstw(Q^2)$ to $\pm 0.3\%$ with little theoretical uncertainty, given the recent advances in the 
evaluation of $\gamma\zzero$ box diagrams. We will soon see if the result presents any surprises.

The tremendous technical advances in experimental methods as well as the theoretical advances in precision
calculations have set the stage for very low $Q^2$ $ee$ and $ep$ asymmetry measurements that will aim for 
unprecedented $\pm 0.1\%$ $\sstw(Q\approx 0)$ sensitivity, allowing them to be competitive with the best
$\zzero$ pole studies. At that level, $Z'$ models and contact interactions are probed at mass scales in the 
1--20 TeV range, the running of $\sstw(Q^2)$ is precisely verified or perhaps dramatically new phenomena
are uncovered. For example, low $Q^2$ measurements may unveil ``Dark Parity Violation" due to a very light,
weakly coupled $Z_d$ boson from the dark matter sector. If found, such a discovery would revolutionize the
focus of low energy parity violation experiments and provide strong motivation for other challenging 
low and high energy
experiments designed to shed new light on ``Dark Physics". 

Surprises advance science, but they are only possible 
if we push the boundaries of our abilities.

\section*{Acknowledgment} 
We thank J. Erler, Y. Li and M. Ramsey-Musolf for useful discussion and input on figures.
This work has been funded in part by the United States Department of Energy under grant contract numbers
 DE-FG02-88R40415-A018 (KK), DE-AC02-98CH10886 (WJM), DE-FG02-84ER40146 (PAS), and
 by the U.S. National Science Foundation under grant NSF-PHY- 0705682 (SM). WJM acknowledges partial support as a Fellow in the Gutenberg Research College.

\bibliographystyle{arnuke_revised.bst}
\bibliography{lowesin2theta_mods.bib}
\end{document}